\documentstyle[prl,aps,preprint,epsf,floats]{revtex}

\newcommand{\sh}{\!\!\! /}

\begin{document}
\setcounter{page}{0}
\def\footnoterule{\kern-3pt \hrule width\hsize \kern3pt}

\title{Conformal Symmetry and the Three Point Function for the Gravitational 
Axial Anomaly}

\author{Jiannis Pachos\footnote{E-mail address: {\tt pachos@ctp.mit.edu}} 
and Ricardo Schiappa\footnote{E-mail address: {\tt ricardos@ctp.mit.edu}} }

\bigskip

\bigskip

\bigskip

\address{
Center for Theoretical Physics and Department of Physics\\
Massachusetts Institute of Technology, 77 Massachusetts Ave. \\
Cambridge, MA 02139, U.S.A. \\
}

\medskip

\date{{\sc MIT-CTP$\sharp$2760},~ {\tt hep-th/9807128},~ June 1998}
\maketitle

\thispagestyle{empty}

\begin{abstract}
This work presents a first study of a radiative calculation for the 
gravitational axial anomaly in the massless Abelian Higgs model. The 
two loop contribution to the anomalous correlation function of one 
axial current and two energy-momentum tensors, 
$\langle A_\alpha(z) T_{\mu\nu}(y) T_{\rho\sigma}(x) \rangle$, is 
computed at an order that involves only internal matter fields. Conformal 
properties of massless field theories are used in order to perform the 
Feynman diagram calculations in the coordinate space representation. The 
two loop contribution is found not to vanish, due to the presence of two 
independent tensor structures in the anomalous correlator.
\end{abstract}

\bigskip


\medskip


\vspace*{\fill}

\pacs{xxxxx}

\section{Introduction and Discussion}

Perturbation theory anomalies have been known for a long time, starting 
with the work by Bell and Jackiw \cite{bell} and by Adler \cite{adler} 
concerning anomalies in gauge theories. The fact that there are no radiative 
corrections to the one loop result for the anomaly has been countlessly 
proven or brought into question since the early work of Adler and 
Bardeen \cite{adbard}. For this reason, explicit calculations of possible 
radiative corrections to the one loop anomaly are of particular interest.

Using a method proposed by Baker and Johnson \cite{kj1}, Erlich and Freedman 
recently performed such an explicit calculation for the two loop 
contribution of the anomalous correlation function 
$\langle A_\mu (x) A_\nu (y) A_\rho (z) \rangle$ of three chiral currents, 
in the Abelian Higgs model and in the Standard Model \cite{JF}. In here, we 
wish to extend such calculation to the case of a gravitational background.

The Adler-Bell-Jackiw anomaly concerns the divergence of the axial 
current in a gauge field background. The calculation of the divergence of 
the axial current in a gravitational field background was later performed 
by Delbourgo and Salam \cite{del1}, Eguchi and Freund \cite{eguchi} and 
Delbourgo \cite{del2}. As in the ABJ case, these authors found an anomaly 
associated to the conservation of the axial current, the gravitational 
axial anomaly. Later, Alvarez-Gaum\'e and Witten showed the 
significance of gravitational anomalies for a wide variety of physical 
applications \cite{witt}.

The question of absence of radiative corrections to the one loop result 
obtained for the gravitational axial anomaly is an issue 
not as well established as it is in the gauge theory case. This 
is the reason why we proceed to perform an explicit two loop calculation, 
adopting the spirit in \cite{JF}. However, calculating the two loop 
contribution to the gravitational axial anomaly is a much longer 
task than to do so for the gauge axial anomaly. In this paper we shall 
address the first part of the computation, by calculating the abnormal 
parity part of the three point function involving one axial vector and 
two energy-momentum tensors at a specific two loop order in the Abelian 
Higgs model. The reason we choose to work in this model is due to the 
recent interest arising from the gauge anomaly case in \cite{JF}, and 
also due to the fact that this model is a simplified version of the 
Standard Model. In order to set notation, the anomalous correlator we 
shall be dealing with is:
$$
\langle A_\alpha(z) T_{\mu\nu}(y) T_{\rho\sigma}(x) \rangle, 
\eqno(1.1)
$$
where $A_\alpha$ is the axial current and $T_{\mu\nu}$ the energy-momentum 
tensor.

The method of calculation \cite{kj1,JF} is based on conformal properties 
of massless field theories, and also involves ideas from the coordinate 
space method of differential regularization due to Freedman, Johnson and 
Latorre \cite{kj2}. In particular, the correlator (1.1) will be directly 
calculated in Euclidean position space and a change of variables suggested 
by the conformal properties of the correlator will be used in order to 
simplify the internal integrations. The order in two loops we shall 
be working involves no internal photons, but only internal matter 
fields (the scalar and spinor fields in the Abelian Higgs model). However, 
in this case diagrams containing vertex and self-energy corrections will 
require a regularization scale. To handle this technicality we shall 
introduce photons in our calculation, as there is a unique choice of gauge 
fixing parameter (in the photon propagator) which makes both the self-energies 
and vertex corrections finite. These ``finite gauge photons'' are merely
a technical tool employed in the calculation.

The use of conformal symmetry to construct three point functions is 
well established. Of particular interest to us is the work by Schreier 
\cite{schreier}, where three point functions invariant under conformal 
transformations were constructed. For the case of one axial and two vector 
currents, it was shown that there is a unique conformal tensor present in the 
three point function. More recently, Osborn and Petkos \cite{osborn1} and 
Erdmenger and Osborn \cite{osborn2} have used conformal invariance to 
compute several three point functions involving the energy-momentum tensor. 
However, the case of one axial current and two energy-momentum tensors 
was not considered.

What we find in here is that, even though at one loop there is only one 
conformal tensor present in the correlator (1.1) -- the one that leads to 
the contraction of the Riemann tensor with its dual in the expression for 
the anomaly --, at two loops there are two independent conformal tensors 
present in the correlator. This is unlike the gauge axial anomaly case 
where the only possible tensor is the one that leads to the field strength 
contracted with its dual in the anomaly equation. Precisely because 
of the presence of these two tensors in the two loop result for the 
three point function, this correlator does not vanish. Again, this is 
unlike the gauge axial anomaly case \cite{JF}. 

The two linearly independent conformal tensors present in the anomalous 
correlator are the ones in expressions (3.6) and (3.7) below (where the 
notation is explained in the paragraphs leading up to these formulas). One 
thing we would like to stress is that {\it every} diagram relevant for our 
calculation is either a multiple of one of these tensors, or a linear 
combination of them both.

Two comments are in order. First, the existence of two independent tensors
in the two loop correlator could seem to indicate the existence of a 
radiative correction to the anomaly. On the other hand, the fact that the 
correlator does not vanish at two loops does not mean that its divergence 
(the anomaly) does not vanish at two loops.

Another point of interest is to follow \cite{kj2} and study the differential 
regularization of the one loop triangle diagram associated to the 
gravitational axial anomaly, Figure 1(a). This is done in the Appendix.
What one finds is that differential regulation entails the introduction 
of several different mass scales. Renormalization or symmetry conditions 
may then be used to determine the ratios of these mass scales. In the 
gauge axial anomaly case it was found that there is only one mass ratio 
\cite{kj2}. In this gravitational axial anomaly case, we have shown in the 
Appendix that there is more than one mass ratio. This multiplicity of the 
mass ratios introduces new parameters that could be able to cancel all 
potential (new) anomalies. Apart from presenting part of these different 
scales we shall not proceed with their study. Here, we shall only restrict to 
the calculation of the correlation function, which by itself consists a 
lengthy project. Extracting the two loop contribution to the gravitational 
axial anomaly from our three point function is a question we hope to report 
about in the near future. 

The structure of this paper is as follows. In section 2 we present the 
massless Abelian Higgs model, as well as a review of the basic ideas 
involved in the method of calculation we use. This includes the calculation 
of the one loop triangle diagram. Then, in section 3 we perform our 
two loop calculation, with emphasis on rigorous details. The many 
contributing diagrams are organized into separate groups, and then 
analyzed one at a time.

\section{The Abelian Higgs Model and Conformal Symmetry}

We shall start by presenting the massless Abelian Higgs model. In four 
dimensional Euclidean space, its action is given by:
$$
S=\int d^4x \, \Bigl\{ 
{1\over4}F_{\mu\nu}F_{\mu\nu}+(D_\mu\phi)^\dagger D_\mu\phi+
{\bar{\psi}}\gamma_\mu D_\mu\psi-f{\bar{\psi}}(L\phi+R\phi^\dagger)\psi-
{\lambda\over4}(\phi^\dagger\phi)^2 \Bigr\}, 	\eqno(2.1)
$$
where we have used $L={1\over2}(1-\gamma_5)$ and $R={1\over2}
(1+\gamma_5)$, with $\gamma_5=\gamma_1\gamma_2\gamma_3\gamma_4$. The 
covariant derivatives are:
$$
D_\mu\phi=(\partial_\mu+ig{\cal A}_\mu)\phi,
$$
$$
D_\mu\psi=(\partial_\mu+{1\over2}ig{\cal A}_\mu\gamma_5)\psi,	\eqno(2.2)
$$
so that the theory is parity conserving with pure axial gauge coupling.

Next we introduce a background (external) gravitational field, in order 
to properly define the energy-momentum tensors associated to the 
scalar and spinor matter degrees of freedom. A simple way to do this is 
to couple our model to gravity, so that a spacetime metric $g_{\mu \nu}(x)$ is 
naturally introduced in the Lagrangian as a field variable. Then we can 
obtain the energy-momentum tensor by varying the Lagrangian with respect 
to the metric $g_{\mu \nu}(x)$ as $T_{\mu \nu}(x) = 2 {\delta\over\delta 
g^{\mu\nu}(x)} \int d^4x\,\sqrt{-g}\,{\cal L}$, where $T_{\mu \nu}(x)$ is 
manifestly symmetric. In addition we have to ensure that it is conserved 
and traceless, obtaining finally for the fermion field,
$$
T_{\mu\nu}^{\bf F}={\bar{\psi}}\,\gamma_{(\mu}\partial_{\nu)}\,\psi, 
\eqno(2.3)
$$
and for the boson field,
$$
T_{\mu\nu}^{\bf B}={2\over3}\Bigl\{\partial_\mu{\phi^\dagger}\,
\partial_\nu\phi+\partial_\nu{\phi^\dagger}\,\partial_\mu\phi-
{1\over2}\delta_{\mu\nu}\partial_\alpha{\phi^\dagger}\,\partial_\alpha\phi-
{1\over2}(\phi\,\partial_\mu\partial_\nu{\phi^\dagger}+{\phi^\dagger}\,
\partial_\mu\partial_\nu\phi)\Bigr\},	\eqno(2.4)
$$
where $(\mu\nu)\equiv\mu\nu+\nu\mu$.

One should observe that in the two loop calculation we are interested in 
computing the order ${\cal O}(gf^2k^2)$ correction to the correlator, 
where $g$ is the gauge coupling, $f$ the scalar-spinor coupling, and 
$k$ the gravitational coupling. This means that there are no internal 
photons in the associated diagrams, as these would be of order 
${\cal O}(g^3k^2)$ -- we shall only need photons as the external 
axial current, and in order to handle some of the potential 
divergences in the calculation (see section 3). 
This is why in (2.3) and (2.4) the scalar and spinor matter degrees of 
freedom are decoupled from the gauge field.

Conformal symmetry plays a central role in our calculations, as it 
motivates a change of variables that simplifies the two 
loop integrations. Due to the absence of any scale, 
our model is conformal invariant. The conformal group of Euclidean field 
theory is $O(5,1)$ \cite{osborn1}.
All transformations which are continuously 
connected to the identity are obtained via a combination of rotations 
and translations with the basic conformal inversion,
$$
x_\mu={x'_\mu\over{x'}^2},
$$
$$
{\partial x_\mu \over \partial x'_\nu}=x^2
(\delta_{\mu\nu}-{2x_\mu x_\nu \over x^2}) \equiv
x^2 J_{\mu\nu}(x).	\eqno(2.5)
$$
The Jacobian tensor, $J_{\mu\nu}(x)$, which is an improper orthogonal matrix 
satisfying $J_{\mu\nu}(x)=J_{\mu\nu}(x')$, will play a useful role in the 
calculation of the coordinate space Feynman diagrams.

The action (2.1) is invariant under conformal inversions, as \cite{JF}:
$$
\phi(x) \rightarrow \phi'(x)={x'}^2\phi(x'),
$$
$$
\psi(x) \rightarrow \psi'(x)={x'}^2\gamma_5{x\sh}'\psi(x'),
$$
$$
{\cal A}_\mu(x) \rightarrow {{\cal A}'}_\mu(x)=-{x'}^2J_{\mu\nu}(x')
{\cal A}_\nu(x'),	\eqno(2.6)
$$
while also the following relations hold,
$$
d^4x={d^4x' \over {x'}^8} \qquad {\rm and} \qquad 
{x\sh}'\gamma_\mu{x\sh}'=-{x'}^2J_{\mu\nu}(x')\gamma_\nu. 
\eqno(2.7)
$$

In order to use conformal properties to simplify the two loop Feynman 
integrals, one should expect that the relevant Feynman rules 
will consist of vertex factors and propagators with simple inversion 
properties. In particular for the scalar and spinor propagators we have,
$$
\Delta(x-y)={1 \over 4\pi^2}{1 \over (x-y)^2}={1 \over 4\pi^2}
{{x'}^2{y'}^2 \over (x'-y')^2},
$$
$$
S(x-y)=-{\partial\sh}\Delta(x-y)={1 \over 2\pi^2}{{x\sh}-{y\sh}
\over (x-y)^4}=-{1\over 2\pi^2}{x'}^2{y'}^2 {x\sh}'
{{x\sh}'-{y\sh}' \over (x'-y')^4}{y\sh}'.	\eqno(2.8)
$$
The vertex rules, read from the action (2.1), are:

\begin{figure}[ht]
  \centerline{
    \put(120,185){$=-fL,$}
    \put(320,185){$=-fR,$}
    \put(140,105){$=-$ $\large{{1\over2}}$ $ig\,\gamma_\alpha\gamma_5\,
\delta^4(z-z_1)\,\delta^4(z-z_2),$}
    \put(140,28){$=ig$ $\large{({\partial\over\partial z_2^\alpha}}$ $-$ 
$\large{{\partial\over\partial z_1^\alpha})}$ 
$\delta^4(z-z_1)\,\delta^4(z-z_2),$}
    \put(420,185){(2.9)}
    \put(414,105){(2.10)}
    \put(414,28){(2.11)}
    \epsfxsize=6in
    \epsfysize=3.5in
    \epsffile{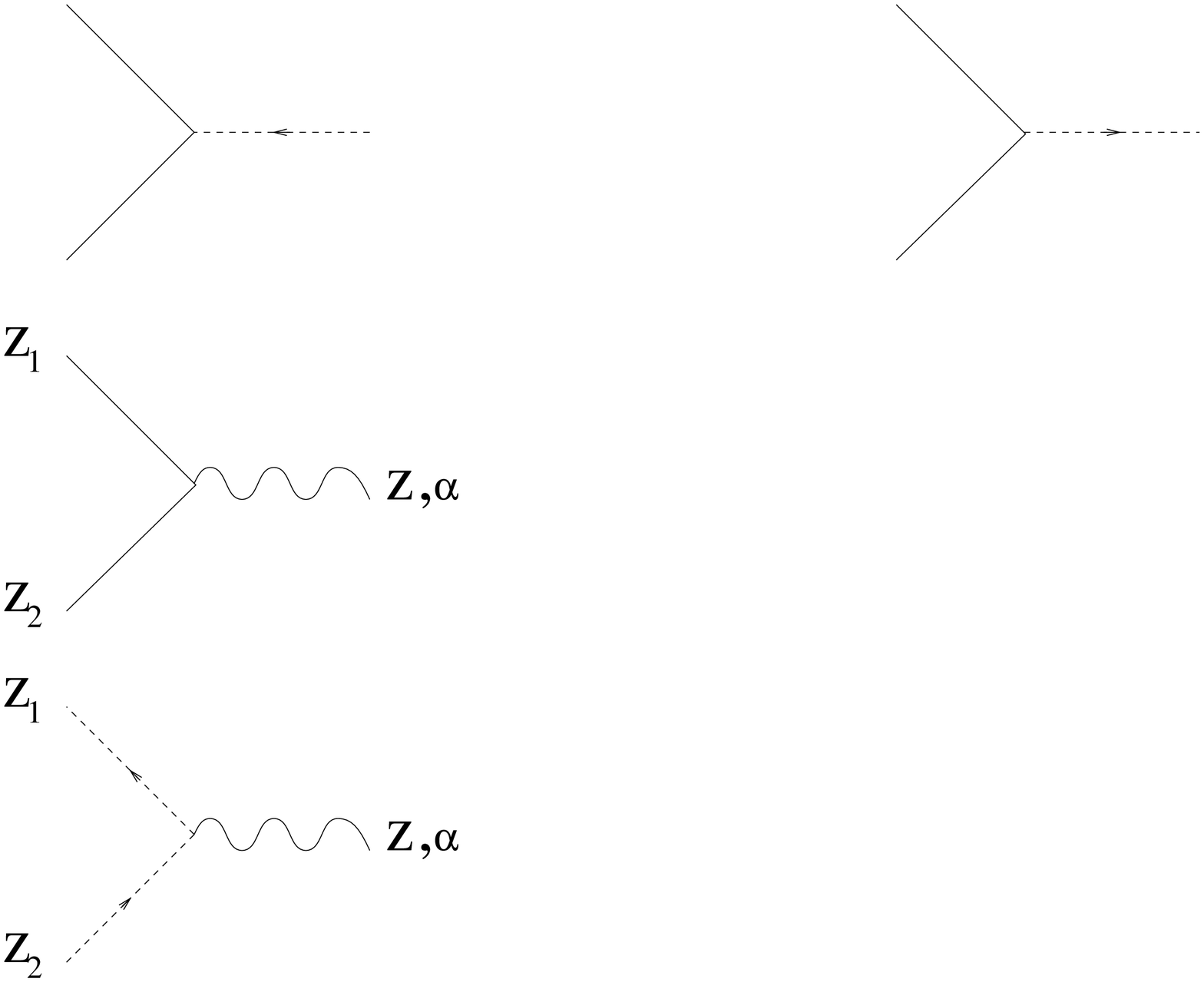}
             }
\end{figure}

\noindent
where solid lines are fermions, dashed lines are scalars and wavy lines 
are gauge fields.

In addition the energy-momentum tensor insertions (2.3) and (2.4) yield 
the following vertices:

\begin{figure}[ht]
  \centerline{
    \put(140,102){$=k\gamma_{(\mu}\delta_{\nu)\alpha}$ $\large{({\partial\over
\partial z_2^\alpha}}$ $-$ $\large{{\partial\over\partial z_1^\alpha})}$ 
$\delta^4(z-z_1)\,\delta^4(z-z_2),$}
    \put(140,31){$=$ $\large{{2\over3}}$ $k$ 
$\large{({\partial\over\partial z_2^{(\mu}}{\partial\over\partial 
z_1^{\nu)}}}$ $-$ $\large{{1\over2}}$ $\delta_{\mu\nu}$ $\large{{\partial
\over\partial z_2^\alpha}{\partial\over\partial z_1^\alpha}}$ $-$ 
$\large{{1\over2}\,({\partial^2\over\partial z_2^\mu\partial z_2^\nu}}$ $+$ 
$\large{{\partial^2\over\partial z_1^\mu\partial z_1^\nu}))}\cdot$}
    \put(240,0){$\cdot\delta^4(z-z_1)\,\delta^4(z-z_2),$}
    \put(430,102){(2.12)}
    \put(430,0){(2.13)}
    \epsfxsize=6.5in
    \epsfysize=2in
    \epsffile{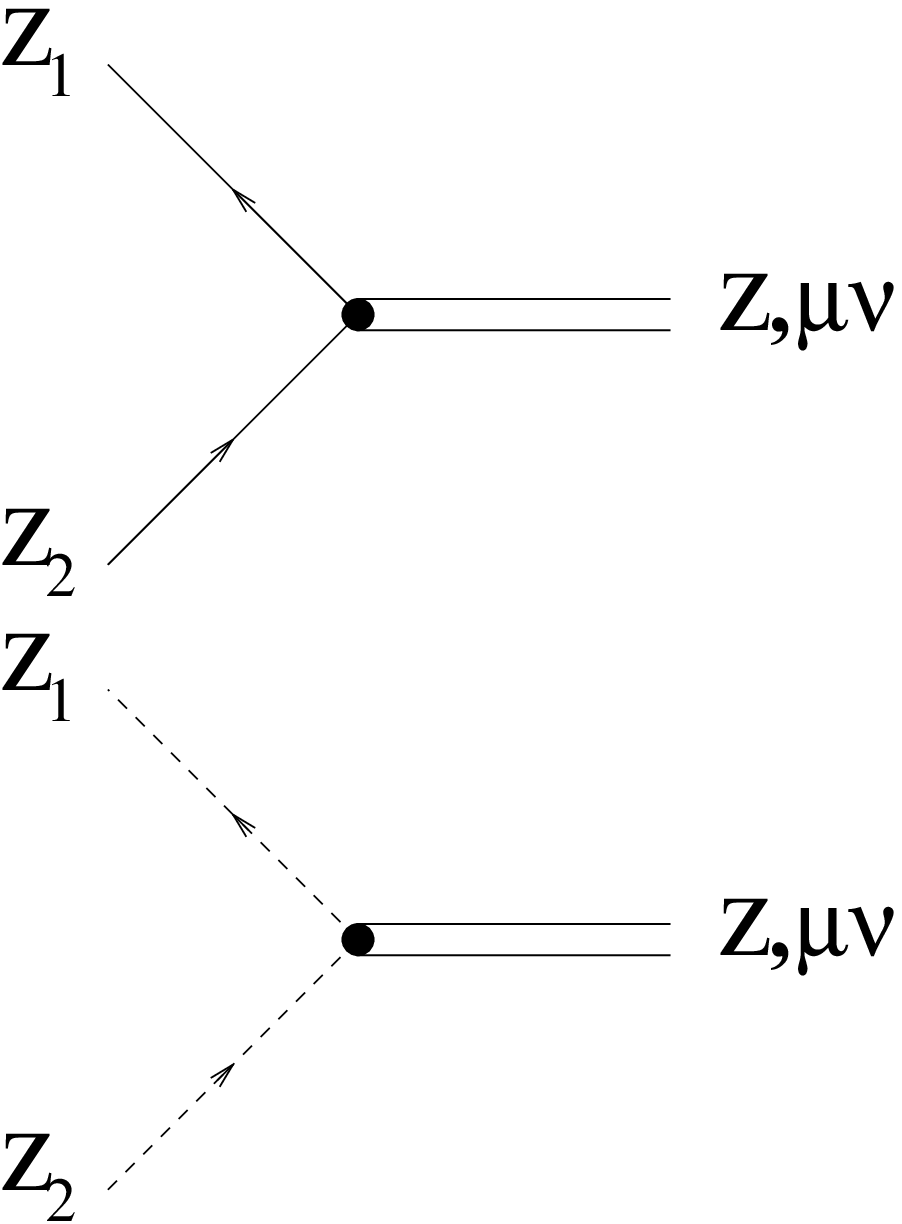}
             }
\end{figure}

\noindent
where the double solid lines represent gravitons.

Let us analyze the conformal properties of the graviton vertices. In 
order to do that, we attach the vertices (2.12) and (2.13) to 
scalar and spinor legs, and use (2.5), (2.7) and (2.8) to obtain,
$$
S(v-x)\,\gamma_{(\mu}\delta_{\nu)\alpha}\,(
{\overrightarrow{\partial}\over\partial x_\alpha}-
{\overleftarrow{\partial}\over\partial x_\alpha})\,S(x-u)=
$$
$$
=-{x'}^8\,J_{\bar{\mu}(\mu}(x')\,J_{\nu)\bar{\nu}}(x')\,
\Bigl\{\,
{v'}^2{v\sh}'\,S(v'-x')\,\gamma_{\bar{\mu}}\,(
{\overrightarrow{\partial}\over\partial x'_{\bar{\nu}}}-
{\overleftarrow{\partial}\over\partial x'_{\bar{\nu}}})\,
S(x'-u')\,{u'}^2{u\sh}'
\,\Bigr\},	\eqno(2.14)
$$
for the fermionic vertex, where the derivatives {\it only} act inside 
the curly brackets. Likewise,
$$
\Delta(v-x)\,(
{\overleftarrow{\partial}\over\partial x_{(\mu}}
{\overrightarrow{\partial}\over\partial x_{\nu)}}-{1\over2}\delta_{\mu\nu}\,
{\overleftarrow{\partial}\over\partial x_\alpha}
{\overrightarrow{\partial}\over\partial x_\alpha}-{1\over2}\,(
{\overleftarrow{\partial^2}\over\partial x_\mu\partial x_\nu}+
{\overrightarrow{\partial^2}\over\partial x_\mu\partial x_\nu}))\,
\Delta(x-u)={x'}^8\,J_{\bar{\mu}\mu}(x')\,J_{\nu\bar{\nu}}(x')\cdot
$$
$$
\cdot\Bigl\{ {v'}^2\Delta(v'-x')\,(
{\overleftarrow{\partial}\over\partial x'_{(\bar{\mu}}}
{\overrightarrow{\partial}\over\partial x'_{\bar{\nu})}}-
{1\over2}\delta_{\bar{\mu}\bar{\nu}}\,
{\overleftarrow{\partial}\over\partial x'_\alpha}
{\overrightarrow{\partial}\over\partial x'_\alpha}-{1\over2}\,(
{\overleftarrow{\partial^2}\over\partial x'_{\bar{\mu}}
\partial x'_{\bar{\nu}}}+
{\overrightarrow{\partial^2}\over\partial x'_{\bar{\mu}}
\partial x'_{\bar{\nu}}}))\,
\Delta(x'-u'){u'}^2\Bigr\},		\eqno(2.15)
$$
for the bosonic vertex, where once again the derivatives only act 
inside the curly brackets.

As an illustration of these coordinate space propagators and vertex 
rules, we shall now look at the one loop triangle diagram and 
perform the conformal inversion on the amplitude's tensor structure. 
The relevant one loop triangle diagram is depicted in Figure 1(a) and 
its amplitude, $B_{\alpha, \mu\nu, \rho\sigma}(z,y,x)$, can be computed 
using the previous rules to be:
$$
B_{\alpha, \mu\nu, \rho\sigma}(z,y,x)=
$$
$$
={1\over2}igk^2\,{\bf Tr}\,\gamma_\alpha\gamma_5\,S(z-y)\,
\gamma_{(\mu}\delta_{\nu)\beta}\,(
{\overrightarrow{\partial}\over\partial y_\beta}-
{\overleftarrow{\partial}\over\partial y_\beta})\,S(y-x)\,
\gamma_{(\rho}\delta_{\sigma)\pi}\,(
{\overrightarrow{\partial}\over\partial x_\pi}-
{\overleftarrow{\partial}\over\partial x_\pi})\,S(x-z). 
\eqno(2.16)
$$
Due to translation symmetry we are free to set $z=0$, 
while we refer the remaining external points $x$ and $y$ to 
their inverted images (2.5). Although this transformation may seem 
{\it ad hoc} at this stage, it will later simplify the calculation of the 
two loop diagrams \cite{JF}. The result we obtain is,
$$
B_{\alpha, \mu\nu, \rho\sigma}(0,y,x)
=-{igk^2\over8\pi^4}{y'}^8{x'}^8\,
J_{\bar{\mu}(\mu}(y')J_{\nu)\bar{\nu}}(y')J_{\bar{\rho}(\rho}(x')
J_{\sigma)\bar{\sigma}}(x')\,{\bf Tr}\,\gamma_\alpha\gamma_5
\gamma_{\bar{\mu}}\,
{\partial^2\over\partial y'_{\bar{\nu}}\partial x'_{\bar{\sigma}}}
\,S(y'-x')\,
\gamma_{\bar{\rho}}.	\eqno(2.17)
$$
Taking the fermionic trace one finally gets,
$$
B_{\alpha, \mu\nu, \rho\sigma}(0,y,x)\,
=\,{igk^2\over4\pi^6}\,{y'}^8{x'}^8\,
J_{\bar{\mu}(\mu}(y')\,J_{\nu)\bar{\nu}}(y')\,J_{\bar{\rho}(\rho}(x')
\,J_{\sigma)\bar{\sigma}}(x')\,
\varepsilon_{\alpha\kappa\bar{\mu}\bar{\rho}}\,
{\partial^2\over\partial y'_{\bar{\nu}}\partial x'_{\bar{\sigma}}}
\,{(x'-y')_\kappa\over(x'-y')^4}.	\eqno(2.18)
$$

For separated points (2.16) is fully Bose symmetric and conserved on 
all indices. The expected anomaly is a local violation of the 
conservation Ward identities which arises because the differentiation 
of singular functions is involved \cite{JF}. There are several ways 
to obtain the anomaly in this coordinate space approach 
\cite{kj2,sonoda,JF}. One way \cite{kj2,JF} to do this is to recognize 
that the amplitude (2.16) is too singular at short distances to have a 
well defined Fourier transform. One then regulates, which entails the 
introduction of several independent mass scales. The regulated 
amplitude is well defined, and one can check the Ward identities. Also, 
an important aspect of this coordinate space approach to the axial 
anomaly is that the well defined amplitude (2.16), for separated points, 
determines the fact that there is an anomaly of specific strength \cite{kj2}.

Some more comments about the role of the conformal symmetry in the 
calculation of possible radiative corrections to the anomaly are 
now in order. At first sight this could look as a questionable role, 
after all the introduction of a scale 
to handle the divergences of perturbation 
theory will spoil any expected conformal properties. 
This is true in general, 
but our two loop triangle diagrams for this massless Abelian Higgs 
model are exceptional. Any primitively divergent amplitude is 
exceptional when studied in coordinate space for separated points, 
since the internal integrals converge without regularization \cite{JF}.

As we shall see in the next section, there will be 3 non-planar and 3 planar 
diagrams, which are primitives. Of course there will be many other 
diagrams which contain sub-divergent vertex and self-energy corrections, 
and these require a regularization scale. However, we are dealing with 
pure axial coupling for the fermion. This means that if we introduce 
diagrams containing internal photons, then there is a unique 
choice of gauge-fixing parameter $\Gamma$ which makes the one loop 
self-energy finite \cite{JF}. Moreover, since the vertex and 
self-energy corrections are related by a Ward identity, each vertex 
correction is also finite in this same gauge. In conclusion, choosing 
this finite gauge makes it possible to obtain a finite two loop 
result in our calculation.

\section{The Three Point Function for the Two Loop Gravitational Axial Anomaly}

Let us now proceed to the next loop order in the non-gauge sector, as we are 
interested in computing possible corrections to the gravitational axial 
anomaly at order ${\cal O}(gf^2k^2)$. At this order we have a total 
of 36 diagrams that can possibly contribute. Of these diagrams, 3 are 
non-planar, but they actually only correspond to 2 independent 
calculations due to reflection symmetry. These are depicted in 
Figures 1(b) and 1(c). Then, there are 3 scalar self-energy 
diagrams, and other 3 photon self-energy diagrams (as we shall 
see, some diagrams involving photons are required in order 
to choose the finite gauge and compensate some divergences of the non-gauge
amplitudes). These 6 self-energy diagrams amount to 2 independent 
calculations alone, the ones depicted in Figures 1(d) and 1(e). Then, we 
have 3 axial current insertion vertex corrections, in Figures 1(f), 1(g) 
and 1(h). At the energy-momentum tensor insertion, we also have 
vertex corrections. These are 6 diagrams, amounting to the 3 independent 
calculations in Figures 1(i), 1(j) and 1(k). There are also 6 diagrams 
that identically vanish due to fermionic traces, the ones in Figures 1(l), 
1(m) and 1(n). Associated to the mentioned self-energies there 
are 3 diagrams corresponding to local self-energy renormalizations. 
They amount to 1 independent calculation, Figure 1(o). Also, associated to 
the mentioned vertex corrections at the energy-momentum insertion there 
are 2 diagrams corresponding to local vertex renormalizations. They amount 
to 1 independent calculation, Figure 1(p). So, overall, of the 29 initial 
two loop diagrams in Figure 1, we are left with 12 independent calculations. 
In Figure 2 we have 7 more diagrams, corresponding to 4 independent 
calculations. These diagrams are associated to the finite gauge photons and 
shall be discussed later. We are thus left with an overall number of 16 
independent calculations, out of the initial 36 diagrams. Let us see how to 
perform such calculations, one at a time.

\subsection{Diagrams in Figures 1(b) and 1(c)}

We begin with the non-planar diagram depicted in Figure 1(b), which we 
shall denote by $N^{(1)}_{\alpha, \mu\nu, \rho\sigma}(z,y,x)$.
This amplitude is conformal covariant since no issues of subdivergences 
and gauge choice arise. The idea \cite{kj1} is to use the inversion, 
$u_\alpha={u'}_\alpha/{u'}^2$ and $v_\alpha={v'}_\alpha/{v'}^2$, as a 
change of variables in the internal integrals. In order to use the 
simple conformal properties of the propagators (2.8) we must also refer 
the external points to their inverted images (2.5), as was done 
in (2.14), (2.15), and in (2.17), (2.18). If in succession 
we use the translation symmetry to place one point at the origin, 
say $z=0$, then the propagators attached to that point drop out of 
the integral, because the inverted point is now at $\infty$, and the 
integrals simplify.

After summing over both directions of Higgs field propagation, and setting 
$z=0$, the amplitude for $N^{(1)}_{\alpha, \mu\nu, \rho\sigma}(0,y,x)$ 
is written as,
$$
N^{(1)}_{\alpha, \mu\nu, \rho\sigma}(0,y,x)=
-{igf^2k^2\over8\pi^4}\int d^4u\,d^4v\,
({u_\alpha\over u^4v^2}-{v_\alpha\over v^4u^2})\cdot
$$
$$
\cdot{\bf Tr}\,\gamma_5\,
S(v-x)\gamma_{(\rho}\delta_{\sigma)\beta}\,(
{\overrightarrow{\partial}\over\partial x_\beta}-
{\overleftarrow{\partial}\over\partial x_\beta})\,S(x-u)S(u-y)
\gamma_{(\mu}\delta_{\nu)\pi}\,(
{\overrightarrow{\partial}\over\partial y_\pi}-
{\overleftarrow{\partial}\over\partial y_\pi})\,S(y-v).	\eqno(3.1)
$$
The change of variables previously outlined can be performed with the 
help of (2.7), (2.8), and the Higgs current transformation,
$$
{u_\alpha\over u^4v^2}-{v_\alpha\over v^4u^2}=
{v'}^2{u'}^2\,({u'}_\alpha-{v'}_\alpha).	\eqno(3.2)
$$
The spinor propagator side factors ${x\sh}'$, ${v\sh}'$, etc., 
collapse within the trace, and the Jacobian $({u'}{v'})^{-8}$ cancels 
with factors in the numerator. Performing the algebra we obtain,
$$
N^{(1)}_{\alpha, \mu\nu, \rho\sigma}(0,y,x)=
-{igf^2k^2\over8\pi^4}\,{y'}^8{x'}^8\,
J_{\bar{\mu}(\mu}(y')\,J_{\nu)\bar{\nu}}(y')\,J_{\bar{\rho}(\rho}(x')
\,J_{\sigma)\bar{\sigma}}(x')
\int d^4{u'}\,d^4{v'}\,(v'-u')_\alpha\cdot
$$
$$
\cdot\Bigl\{ {\bf Tr}\,\gamma_5\,
S(v'-x')\gamma_{\bar{\rho}}\,(
{\overrightarrow{\partial}\over\partial {x'}_{\bar{\sigma}}}-
{\overleftarrow{\partial}\over\partial {x'}_{\bar{\sigma}}})\,
S(x'-u')S(u'-y')\gamma_{\bar{\mu}}\,(
{\overrightarrow{\partial}\over\partial {y'}_{\bar{\nu}}}-
{\overleftarrow{\partial}\over\partial {y'}_{\bar{\nu}}})\,
S(y'-v')\Bigr\},	\eqno(3.3)
$$
where the derivatives are acting only inside the curly brackets.

We see that we obtain the expected transformation 
factors for the energy-momentum tensors at $x$ and $y$ times an 
integral in which $u'$ and $v'$ each appear in only two 
denominators. These convolution integrals can be done in 
several different ways. The final relevant formulas are listed 
in the Appendix. We begin by using the trace properties to move 
the $S(y'-v')$ propagator in (3.3) close to the $S(v'-x')$ propagator. 
The differential operators are kept fixed, with the understanding that 
now the $y'$ derivative that seems to be acting on nothing is actually 
acting on the propagator $S(y'-v')$ which is now sitting on the left. As 
usual, all derivatives act only inside curly brackets. We can perform the 
integrations without the need to make the differentiations first as 
the integration variables are well separated from the differentiation ones.
Expand the product with $(v'-u')_\alpha$, and we are led to the 
following result:
$$
\int d^4{u'}d^4{v'}(v'-u')_\alpha
{\bf Tr}\gamma_5
S(v'-x')\gamma_{\bar{\rho}}(
{\overrightarrow{\partial}\over\partial {x'}_{\bar{\sigma}}}-
{\overleftarrow{\partial}\over\partial {x'}_{\bar{\sigma}}})
S(x'-u')S(u'-y')\gamma_{\bar{\mu}}(
{\overrightarrow{\partial}\over\partial {y'}_{\bar{\nu}}}-
{\overleftarrow{\partial}\over\partial {y'}_{\bar{\nu}}})
S(y'-v')=
$$
$$
-{\bf Tr}\gamma_5
\int d^4{v'}v'_\alpha S(v'-y')S(v'-x')\,\gamma_{\bar{\rho}}(
{\overrightarrow{\partial}\over\partial {x'}_{\bar{\sigma}}}-
{\overleftarrow{\partial}\over\partial {x'}_{\bar{\sigma}}})
\int d^4{u'}S(u'-x')S(u'-y')\,\gamma_{\bar{\mu}}(
{\overrightarrow{\partial}\over\partial {y'}_{\bar{\nu}}}-
{\overleftarrow{\partial}\over\partial {y'}_{\bar{\nu}}})
$$
$$
+{\bf Tr}\gamma_5
\int d^4{v'}S(v'-y')S(v'-x')\,\gamma_{\bar{\rho}}(
{\overrightarrow{\partial}\over\partial {x'}_{\bar{\sigma}}}-
{\overleftarrow{\partial}\over\partial {x'}_{\bar{\sigma}}})
\int d^4{u'}u'_\alpha S(u'-x')S(u'-y')\,\gamma_{\bar{\mu}}(
{\overrightarrow{\partial}\over\partial {y'}_{\bar{\nu}}}-
{\overleftarrow{\partial}\over\partial {y'}_{\bar{\nu}}}), 
\eqno(3.4)
$$
where the integrals can be directly read off from the Appendix. When 
these results are used and substituted within the trace, one finds 
the final amplitude,
$$
N^{(1)}_{\alpha, \mu\nu, \rho\sigma}(0,y,x)=
-{igf^2k^2\over32\pi^8}\,{y'}^8{x'}^8\,
J_{\bar{\mu}(\mu}(y')\,J_{\nu)\bar{\nu}}(y')\,J_{\bar{\rho}(\rho}(x')
\,J_{\sigma)\bar{\sigma}}(x')\cdot
$$
$$
\cdot\Bigl\{\varepsilon_{\alpha\kappa{\bar{\rho}}{\bar{\mu}}}\,
{1\over(x'-y')^2}\,(
{\overrightarrow{\partial}\over\partial {x'}_{\bar{\sigma}}}-
{\overleftarrow{\partial}\over\partial {x'}_{\bar{\sigma}}})\,
{(x'-y')_\kappa\over(x'-y')^2}\,(
{\overrightarrow{\partial}\over\partial {y'}_{\bar{\nu}}}-
{\overleftarrow{\partial}\over\partial {y'}_{\bar{\nu}}})\Bigr\}, 
\eqno(3.5)
$$
where we should have the ``trace'' attitude for taking derivatives: 
the derivatives {\it only} act inside the curly brackets, and the $y'$ 
derivative that seems to be acting on nothing is actually acting on the 
first $(x'-y')$ term.

We observe that unlike the case in \cite{JF}, this non-planar amplitude 
is {\it not} a numerical multiple of the amplitude for the one loop triangle 
diagram (2.18). This is because the tensor (derivative) structure 
in (3.5) is different from the one in (2.18). To see that one just 
has to explicitly compute both structures, and compare them. For 
the triangle one has:
$$
{\partial^2\over\partial x'_{\bar{\sigma}}\partial y'_{\bar{\nu}}}
\,{\Delta_\kappa\over\Delta^4}\,=
$$
$$
=\,{\partial\over\partial x'_{\bar{\sigma}}}\Bigl({\Delta_\kappa\over 
\Delta^2}\Bigr){\partial\over\partial y'_{\bar{\nu}}}\Bigl({1\over\Delta^2}
\Bigr)\,+\,
{\Delta_\kappa\over\Delta^2}{\partial^2\over\partial x'_{\bar{\sigma}}
\partial y'_{\bar{\nu}}}\Bigl({1\over\Delta^2}\Bigr)\,
+\,{1\over\Delta^2}{\partial^2\over\partial x'_{\bar{\sigma}}
\partial y'_{\bar{\nu}}}\Bigl({\Delta_\kappa\over\Delta^2}\Bigr)\,+\,
{\partial\over\partial x'_{\bar{\sigma}}}\Bigl({1\over\Delta^2}\Bigr)
{\partial\over\partial y'_{\bar{\nu}}}\Bigl({\Delta_\kappa\over\Delta^2}
\Bigr),	\eqno(3.6)
$$
while for the non-planar structure one obtains,
$$
{1\over\Delta^2}\,(
{\overrightarrow{\partial}\over\partial {x'}_{\bar{\sigma}}}-
{\overleftarrow{\partial}\over\partial {x'}_{\bar{\sigma}}})\,
{\Delta_\kappa\over\Delta^2}\,(
{\overrightarrow{\partial}\over\partial {y'}_{\bar{\nu}}}-
{\overleftarrow{\partial}\over\partial {y'}_{\bar{\nu}}})\,=
$$
$$
=\,{\partial\over\partial x'_{\bar{\sigma}}}\Bigl({\Delta_\kappa\over\Delta^2}
\Bigr){\partial\over\partial y'_{\bar{\nu}}}\Bigl({1\over \Delta^2}\Bigr)\,-
\,{\Delta_\kappa\over\Delta^2}{\partial^2\over\partial x'_{\bar{\sigma}}
\partial y'_{\bar{\nu}}}\Bigl({1\over\Delta^2}\Bigr)\,
-\,{1\over\Delta^2}{\partial^2\over\partial x'_{\bar{\sigma}}
\partial y'_{\bar{\nu}}}\Bigl({\Delta_\kappa\over\Delta^2}\Bigr)\,+\,
{\partial\over\partial x'_{\bar{\sigma}}}\Bigl({1\over\Delta^2}\Bigr)
{\partial\over\partial y'_{\bar{\nu}}}\Bigl({\Delta_\kappa\over\Delta^2}
\Bigr),	\eqno(3.7)
$$
where we defined $\Delta\equiv(x'-y')$. The reason such difference 
can happen is that while in \cite{JF} there is a unique conformal tensor 
structure for the correlator of three axial vector currents, in here we 
have two conformal tensor structures due to the higher dimensionality of 
the correlator of the one axial vector current and the two energy-momentum 
tensors. Also, observe that both these structures (3.6) and (3.7) are to be 
understood as always attached to the appropriate factors of $J_{\mu\nu}(y')$, 
$J_{\rho\sigma}(x')$ and the appropriate powers of $y'$, $x'$. Moreover the 
diagrams that give rise to them obey conservation equations for the 
energy-momentum tensor insertions. (3.6) is associated to the one loop diagram 
in (2.18). It can easily be proved that the conservation equation is obeyed, 
a standard result from \cite{del1,eguchi,del2} (and also form \cite{kj2} once 
we are aware of the relation (A.6) from the Appendix). (3.7) is associated 
to the two loop diagram in (3.5), and one can also explicitly check the 
conservation law for this case. This existence of two conformal structures 
is an extra feature in the discussion of these two loop diagrams, relative 
to the work in \cite{JF}. 

There are 2 more non-planar diagrams, where the scalar vertex is placed 
at $x$ and at $y$. We need to compute them, as they are independent 
of the previous result (we have a scalar-scalar-tensor vertex 
instead of a scalar-scalar-vector vertex, among other different vertices),
but they amount to 1 independent calculation.

So, we proceed with the non-planar diagram in Figure 1(c), denoted in the 
following by $N^{(2)}_{\alpha, \mu\nu, \rho\sigma}(z,y,x)$.
The method of calculation is very similar to the one for the previous 
diagram, and so we shall perform it in here with somewhat less details. 
After summing over both directions of Higgs field propagation, and setting 
$z=0$, the amplitude for $N^{(2)}_{\alpha, \mu\nu, \rho\sigma}(0,y,x)$ 
is written as,
$$
N^{(2)}_{\alpha, \mu\nu, \rho\sigma}(0,y,x)=
-{igf^2k^2\over12\pi^4}\int d^4u\,d^4v\,
{\bf Tr}\,\gamma_5\,{{u\sh}\gamma_\alpha{v\sh}\over u^4v^4}\,
S(v-x)\gamma_{(\rho}\delta_{\sigma)\beta}\,(
{\overrightarrow{\partial}\over\partial x_\beta}-
{\overleftarrow{\partial}\over\partial x_\beta})\,S(x-u)\cdot
$$
$$
\cdot\Delta(u-y)\,(
{\overleftarrow{\partial}\over\partial y_{(\mu}}
{\overrightarrow{\partial}\over\partial y_{\nu)}}-{1\over2}\delta_{\mu\nu}\,
{\overleftarrow{\partial}\over\partial y_\pi}
{\overrightarrow{\partial}\over\partial y_\pi}-{1\over2}\,(
{\overleftarrow{\partial^2}\over\partial y_\mu\partial y_\nu}+
{\overrightarrow{\partial^2}\over\partial y_\mu\partial y_\nu}))\,
\Delta(y-v).	\eqno(3.8)
$$
Performing the conformal inversion is now no harder than it was for 
the previous diagram. The procedure is essentially the same, and if we 
carry out the algebra we obtain,
$$
N^{(2)}_{\alpha, \mu\nu, \rho\sigma}(0,y,x)=
-{igf^2k^2\over24\pi^4}\,{y'}^8{x'}^8\,
J_{\bar{\mu}(\mu}(y')\,J_{\nu)\bar{\nu}}(y')\,J_{\bar{\rho}(\rho}(x')
\,J_{\sigma)\bar{\sigma}}(x')\cdot
$$
$$
\cdot\int d^4{u'}\,d^4{v'}\,\Bigl\{\,{\bf Tr}\,\gamma_5\gamma_\alpha\,
S(v'-x')\gamma_{\bar{\rho}}\,(
{\overrightarrow{\partial}\over\partial {x'}_{\bar{\sigma}}}-
{\overleftarrow{\partial}\over\partial {x'}_{\bar{\sigma}}})\,
S(x'-u')\cdot
$$
$$
\cdot\Delta(u'-y')\,(
{\overleftarrow{\partial}\over\partial {y'}_{({\bar{\mu}}}}
{\overrightarrow{\partial}\over\partial {y'}_{{\bar{\nu})}}}-
{1\over2}\delta_{{\bar{\mu}}{\bar{\nu}}}\,
{\overleftarrow{\partial}\over\partial {y'}_\pi}
{\overrightarrow{\partial}\over\partial {y'}_\pi}-{1\over2}\,(
{\overleftarrow{\partial^2}\over\partial {y'}_{\bar{\mu}}
\partial {y'}_{\bar{\nu}}}+
{\overrightarrow{\partial^2}\over\partial {y'}_{\bar{\mu}}
\partial {y'}_{\bar{\nu}}}))\,
\Delta(y'-v')\,\Bigr\}.	\eqno(3.9)
$$
Once again the expected structure emerges, and all we have to do is to 
perform the integrations. Using the relevant formulas from the Appendix, 
we find the final result as,
$$
N^{(2)}_{\alpha, \mu\nu, \rho\sigma}(0,y,x)=
{igf^2k^2\over384\pi^8}\,{y'}^8{x'}^8\,
J_{\bar{\mu}(\mu}(y')\,J_{\nu)\bar{\nu}}(y')\,J_{\bar{\rho}(\rho}(x')
\,J_{\sigma)\bar{\sigma}}(x')\Bigl\{
\varepsilon_{\alpha\ell{\bar{\rho}}\kappa}\cdot
$$
$$
\cdot
{(x'-y')_\ell\over(x'-y')^2}\,(
{\overrightarrow{\partial}\over\partial {x'}_{\bar{\sigma}}}-
{\overleftarrow{\partial}\over\partial {x'}_{\bar{\sigma}}})\,
{(x'-y')_\kappa\over(x'-y')^2}\,(
{\overleftarrow{\partial}\over\partial {y'}_{({\bar{\mu}}}}
{\overrightarrow{\partial}\over\partial {y'}_{{\bar{\nu})}}}-
{1\over2}\delta_{{\bar{\mu}}{\bar{\nu}}}\,
{\overleftarrow{\partial}\over\partial {y'}_\pi}
{\overrightarrow{\partial}\over\partial {y'}_\pi}-{1\over2}\,(
{\overleftarrow{\partial^2}\over\partial {y'}_{\bar{\mu}}
\partial {y'}_{\bar{\nu}}}+
{\overrightarrow{\partial^2}\over\partial {y'}_{\bar{\mu}}
\partial {y'}_{\bar{\nu}}}))\Bigr\}.	\eqno(3.10)
$$
However, one should note the following. In (3.3) both differential 
operators were first order in the derivatives, but in (3.9) the 
differential operator associated to the vertex (2.13) is actually 
second order. As we expect to have at the end a result similar to 
(3.5) or (2.18), we have to perform one of the derivatives in order 
for both differential operators to become first order. Manipulating 
this result through a somewhat lengthy calculation, one finds:
$$
N^{(2)}_{\alpha, \mu\nu, \rho\sigma}(0,y,x)=
-{igf^2k^2\over128\pi^8}\,{y'}^8{x'}^8\,
J_{\bar{\mu}(\mu}(y')\,J_{\nu)\bar{\nu}}(y')\,J_{\bar{\rho}(\rho}(x')
\,J_{\sigma)\bar{\sigma}}(x')\cdot
$$
$$
\cdot\Bigl\{\varepsilon_{\alpha\kappa{\bar{\rho}}{\bar{\mu}}}\,
{1\over(x'-y')^2}\,(
{\overrightarrow{\partial}\over\partial {x'}_{\bar{\sigma}}}-
{\overleftarrow{\partial}\over\partial {x'}_{\bar{\sigma}}})\,
{(x'-y')_\kappa\over(x'-y')^2}\,(
{\overrightarrow{\partial}\over\partial {y'}_{\bar{\nu}}}-
{\overleftarrow{\partial}\over\partial {y'}_{\bar{\nu}}})\Bigr\}, 
\eqno(3.11)
$$
where the notation is like in (3.5). One realizes that the 
structure here obtained is the same as in (3.5). So, the 3 non-planar 
diagrams have the same tensor structure, which is different from the 
one associated to the one loop triangle. From this result we immediately 
read the last non-planar diagram, the one with 
the vertex involving the scalar fields and energy-momentum 
tensor located at $x$. All we have to do is to exchange $x$ with $y$ 
and $\mu\nu$ with $\rho\sigma$ in (3.11). This actually does not 
change the amplitude (3.11), so that this third diagram contributes with 
the same amount as its reflection symmetric diagram.

Finally, we can add these 3 diagrams, and obtain the non-planar 
contribution to the two loop correlator. The overall contribution is simply:
$$
N_{\alpha, \mu\nu, \rho\sigma}(0,y,x)\,=\,\sum_{i=1}^3\,
N^{(i)}_{\alpha, \mu\nu, \rho\sigma}(0,y,x)\,=
\,-{3igf^2k^2\over64\pi^8}\,{y'}^8{x'}^8\,
J_{\bar{\mu}(\mu}(y')\,J_{\nu)\bar{\nu}}(y')\,J_{\bar{\rho}(\rho}(x')
\,J_{\sigma)\bar{\sigma}}(x')\cdot
$$
$$
\cdot\Bigl\{\varepsilon_{\alpha\kappa{\bar{\rho}}{\bar{\mu}}}\,
{1\over(x'-y')^2}\,(
{\overrightarrow{\partial}\over\partial {x'}_{\bar{\sigma}}}-
{\overleftarrow{\partial}\over\partial {x'}_{\bar{\sigma}}})\,
{(x'-y')_\kappa\over(x'-y')^2}\,(
{\overrightarrow{\partial}\over\partial {y'}_{\bar{\nu}}}-
{\overleftarrow{\partial}\over\partial {y'}_{\bar{\nu}}})\Bigr\}, 
\eqno(3.12)
$$

\subsection{Diagrams in Figures 1(d), 1(e) and 1(o)}

We now proceed to the self-energy diagrams. These will be the same as in 
the three gauge current case \cite{JF}. We shall see the finite gauge 
mechanism for the one loop self-energies and vertex corrections 
coming about, as it handles certain divergences by choosing a 
gauge where they are zero \cite{kj1,JF}. For this cancelation of 
divergences we have introduced the Abelian field
which can be decoupled at the end by setting its coupling to zero.
Let us see how all that works, by starting with the 
Higgs self-energy diagram in Figure 1(d), 
and the photon self-energy diagram in Figure 1(e). 
These are 3 diagrams in Figure 1(d) (as we can place the self-energy loop 
at any of the 3 sides of the triangle), which amount to 1 independent 
calculation, and other 3 diagrams in Figure 1(e) that again amount to 
1 independent calculation. If we remove the self-energy leg from the 
triangle diagram, and add the Higgs and photon contributions we obtain 
\cite{JF},
$$
\Sigma(v-u)={1\over8\pi^4}\,[f^2+{1\over2}g^2(1-\Gamma)]\,
{{v\sh}-{u\sh}\over(v-u)^6}+a\,{\partial\sh}\,\delta^4(v-u), 
\eqno(3.13)
$$
where $\Gamma$ is the gauge fixing parameter coming from the photon 
propagator \cite{kj1,JF}. In this result, the first term is the part 
of the amplitude which is determined by the Feynman rules read from 
the diagrams. It has a linearly divergent Fourier transform, but the 
crucial point is that this amplitude can be made finite by choosing the 
gauge $\Gamma=1+2f^2/g^2$. It then vanishes for separated points. 
However, there is a possible local term, the second term in (3.13), which 
is left ambiguous by the Feynman rules, and is represented in Figure 1(o). 
The constant $a$ will be determined by the Ward identity \cite{JF}.

In order to proceed with the calculation of this constant using the Ward 
identity, we first need to look at the following vertex 
correction diagrams at the axial current insertion: 
Figure 1(f), $T^{(1)}_{\alpha, \mu\nu, \rho\sigma}(z,y,x)$, 
Figure 1(g), $T^{(2)}_{\alpha, \mu\nu, \rho\sigma}(z,y,x)$, and Figure 1(h), 
$T^{(3)}_{\alpha, \mu\nu, \rho\sigma}(z,y,x)$.
Again, we need a diagram involving photons in order to choose the 
previously introduced finite gauge. Also, these 3 diagrams clearly 
correspond to 3 distinct calculations.

The amplitudes of the 3 vertex correction subgraphs in these diagrams are 
the same as in \cite{JF}. Therefore we already know that each 
contribution has a logarithmic divergent Fourier transform, and that 
the sum of the divergent contributions from these 3 vertex subgraphs is 
proportional to $-2f^2-g^2(1-\Gamma)$, therefore vanishing in the 
same gauge that makes the self-energy finite. Henceforth we shall use 
this gauge.

Let us then proceed with the Ward identity calculation, by summarizing 
the result from \cite{JF}. From the amplitudes for the vertex 
subgraphs in the diagrams $T^{(i)}_{\alpha, \mu\nu, \rho\sigma}(z,y,x)$, 
$i=1,2,3$, we obtain the Ward identity for the theory \cite{JF},
$$
{\partial\over\partial z_\alpha}T_\alpha(z,u,v)=-i{1\over2}g\gamma_5
\,(\delta^4(z-u)-\delta^4(z-v))\,\Sigma(u-v), 	\eqno(3.14)
$$
where $T_\alpha=\sum_{i=1}^3T^i_\alpha$, and $T^i_\alpha$ is the vertex 
subgraph in the diagram $T^{(i)}_{\alpha, \mu\nu, \rho\sigma}$.
The constant $a$ in the self-energy (3.13) can be calculated as in 
\cite{JF} -- where basically one works out the LHS in (3.14) (in the 
finite gauge) in order to find the correct value for (3.13) in the RHS --, 
and the final answer is given by
$$
\Sigma(z)={3\over64\pi^2}\,(f^2-{1\over2}g^2)\,
{\partial\sh}\,\delta^4(z).	\eqno(3.15)
$$
Strictly speaking, one should now proceed to verify that the exact same 
result is obtained from the Ward identity associated with the vertex 
correction diagrams at the energy-momen-\\tum tensor insertions, Figures 1(i), 
1(j) and 1(k). This is in fact true, but for pedagogical reasons we shall 
postpone such a proof for a couple of pages.

It is this result for $\Sigma(v-u)$ which is to be used to evaluate the 
local self-energy renormalization, Figure 1(o), therefore yielding the 
correct value for $a$ in (3.13). 
These again are 3 diagrams that amount to 1 independent calculation as 
in Figures 1(d) and 1(e). As (3.15) is purely local, the integral in $u$ 
and $v$ required for the previous diagram is trivial, simply 
yielding a multiple of the one loop triangle amplitude. The final 
result is that the sum of the self-energy insertion diagrams, Figures 1(d), 
1(e) and 1(o), is a multiple of the one loop amplitude, 
$$
\Sigma'_{\alpha, \mu\nu, \rho\sigma}(z,y,x)={3\over64\pi^2}\,
(f^2-{1\over2}g^2)\,B_{\alpha, \mu\nu, \rho\sigma}(z,y,x), 
\eqno(3.16)
$$
exactly like in \cite{JF} as the internal 
fields are the same. Now recall that 
there is a factor of 3 from the triangular symmetry. There is also a 
factor of 2 for opposite directions of fermion charge flow (such term 
was absent in the non-planar diagrams). Finally, we are interested in 
the ${\cal O}(gf^2k^2)$ corrections, so that the term in $g^2$ in 
(3.16) should be discarded. The overall result for the self-energy 
contribution to the two loop correlator is finally,
$$
\Sigma_{\alpha, \mu\nu, \rho\sigma}(0,y,x)={9f^2\over32\pi^2}\,
B_{\alpha, \mu\nu, \rho\sigma}(0,y,x), 
\eqno(3.17)
$$
where we have set $z=0$ (for coherence with the other diagram calculations).

\subsection{Diagrams in Figures 1(f), 1(g) and 1(h)}

We can now proceed the calculation 
of the vertex correction diagrams at the axial current insertion, 
$T^{(i)}_{\alpha, \mu\nu, \rho\sigma}(z,y,x)$, $i=1,2,3$. As for the 
self-energy, the calculation of these 3 diagrams follows from \cite{JF}. We
shall regard each virtual photon diagram as the sum of two graphs, one 
with the photon propagator in the Landau gauge $\Gamma=1$, and the second 
with inversion covariant pure gauge propagator,
$$
\tilde{\Delta}_{\mu\nu}(u-v)=-{1\over4\pi^2}\,{f^2\over g^2}\,
{1\over(u-v)^2}\,J_{\mu\nu}(u-v).	\eqno(3.18)
$$
The Landau gauge diagrams give order ${\cal O}(g^3k^2)$ contributions to 
the two loop correlator, while the remainder gives an order ${\cal O}
(gf^2k^2)$ contribution which is what we are interested in. Therefore -- 
and similarly to what was done from (3.16) to (3.17) -- we shall discard 
the Landau gauge diagrams from our final result, and only use (3.18) 
for the virtual photon propagator in the finite gauge.

With this in mind we turn to the calculation of the diagrams 
$T^{(i)}_{\alpha, \mu\nu, \rho\sigma}(z,y,x)$. 
The method of calculation is similar to the 
one used for the non-planar diagrams, and so we shall follow it here 
without giving details. After summing over both directions of Higgs field 
propagation, and setting $z=0$, the amplitude for 
$T^{(1)}_{\alpha, \mu\nu, \rho\sigma}(0,y,x)$ is written as,
$$
T^{(1)}_{\alpha, \mu\nu, \rho\sigma}(0,y,x)=
-{igf^2k^2\over8\pi^4}\int d^4u\,d^4v\,
\Delta(v-u)\cdot
$$
$$
\cdot{\bf Tr}\,{{u\sh}\over u^4}\,\gamma_\alpha\gamma_5\,
{{v\sh}\over v^4}\,
S(v-y)\gamma_{(\mu}\delta_{\nu)\beta}\,(
{\overrightarrow{\partial}\over\partial y_\beta}-
{\overleftarrow{\partial}\over\partial y_\beta})\,S(y-x)
\gamma_{(\rho}\delta_{\sigma)\pi}\,(
{\overrightarrow{\partial}\over\partial x_\pi}-
{\overleftarrow{\partial}\over\partial x_\pi})\,S(x-u),	\eqno(3.19)
$$
and performing the conformal inversion we are led to the result,
$$
T^{(1)}_{\alpha, \mu\nu, \rho\sigma}(0,y,x)=
-{igf^2k^2\over8\pi^4}\,{y'}^8{x'}^8\,
J_{\bar{\mu}(\mu}(y')\,J_{\nu)\bar{\nu}}(y')\,J_{\bar{\rho}(\rho}(x')
\,J_{\sigma)\bar{\sigma}}(x')
\int d^4{u'}\,d^4{v'}\,\Delta(v'-u')\cdot
$$
$$
\cdot{\bf Tr}\,\gamma_5\gamma_\alpha\,
S(v'-y')\gamma_{\bar{\mu}}\,(
{\overrightarrow{\partial}\over\partial {y'}_{\bar{\nu}}}-
{\overleftarrow{\partial}\over\partial {y'}_{\bar{\nu}}})\,
S(y'-x')\gamma_{\bar{\rho}}\,(
{\overrightarrow{\partial}\over\partial {x'}_{\bar{\sigma}}}-
{\overleftarrow{\partial}\over\partial {x'}_{\bar{\sigma}}})\,
S(x'-u').	\eqno(3.20)
$$
As usual the expected tensorial structure emerges. We are left with 
the integrations to be performed. However, as we have seen, this 
result is divergent; only when we sum the 3 diagrams 
$T^{(i)}_{\alpha, \mu\nu, \rho\sigma}(z,y,x)$, $i=1,2,3$  
the result will be finite, in the finite gauge. So at this stage we should 
include in the calculation (3.20) the equivalent results 
coming from the diagrams $T^{(2)}_{\alpha, \mu\nu, \rho\sigma}(0,y,x)$ 
and $T^{(3)}_{\alpha, \mu\nu, \rho\sigma}(0,y,x)$ -- where for this last 
one we should use {\it only} the inversion covariant pure gauge 
propagator (3.18). The result of including the 3 diagrams all 
together is to produce an integral of a traceless tensor, which is 
convergent, and can be read from the formulas in the Appendix. Hence we 
can write for the net sum of vertex insertions at point $z$, {\it i.e.}, 
$T^{(1)}_{\alpha, \mu\nu, \rho\sigma}(0,y,x)$ plus 
$T^{(2)}_{\alpha, \mu\nu, \rho\sigma}(0,y,x)$ plus 
$T^{(3)}_{\alpha, \mu\nu, \rho\sigma}(0,y,x)$,
$$
T'_{\alpha, \mu\nu, \rho\sigma}(0,y,x)
={igf^2k^2\over256\pi^8}\,{y'}^8{x'}^8\,
J_{\bar{\mu}(\mu}(y')\,J_{\nu)\bar{\nu}}(y')\,J_{\bar{\rho}(\rho}(x')
\,J_{\sigma)\bar{\sigma}}(x')\,
\varepsilon_{\alpha\kappa\bar{\mu}\bar{\rho}}\,
{\partial^2\over\partial y'_{\bar{\nu}}\partial x'_{\bar{\sigma}}}
\,{(x'-y')_\kappa\over(x'-y')^4},	\eqno(3.21)
$$
which is a multiple of the triangle one loop amplitude (2.18). Recalling 
that there is a factor of 2 for opposite directions of fermion charge 
flow, we can finally write for the contribution of the vertex correction 
diagrams (at the axial current insertion) to the two loop correlator,
$$
T_{\alpha, \mu\nu, \rho\sigma}(0,y,x)={f^2\over32\pi^2}\,
B_{\alpha, \mu\nu, \rho\sigma}(0,y,x). 
\eqno(3.22)
$$

\subsection{Diagrams in Figures 2(a) and 2(b)}

In order to obtain the previous result we had to use finite gauge virtual 
photons, as given by the propagator (3.18). If one has a diagram with an 
internal virtual photon, one should expect a factor of $g$ from each of the 
two internal vertices, and so an overall contribution of order 
${\cal O}(g^3k^2)$. However, if one uses an internal finite gauge virtual 
photon, there is an extra factor of $f^2/g^2$ from the propagator 
(3.18), and we therefore obtain an overall contribution of order 
${\cal O}(gf^2k^2)$, which is the order we are interested in. This means that 
one now has to include all diagrams with one internal finite gauge photon 
(3.18).

In particular we have to include one more vertex in our rules, that 
completes (2.9-13). This vertex can be read from the action (2.1) when 
coupled to gravity, and is the following,

\begin{figure}[ht]
  \centerline{
    \put(145,47){$=$ $-$ $\large{{1\over2}}$ $igk\,(\gamma_{(\mu}
\delta_{\nu)\alpha}-$ $\large{{1\over2}}$ $\delta_{\mu\nu}\gamma_\alpha)\,
\gamma_5\,\delta^4(z-z_1)\,\delta^4(z-z_2)\,\delta^4(z-z_3)$,}
    \put(437,20){(3.23)}
    \epsfxsize=6.5in
    \epsfysize=1.3in
    \epsffile{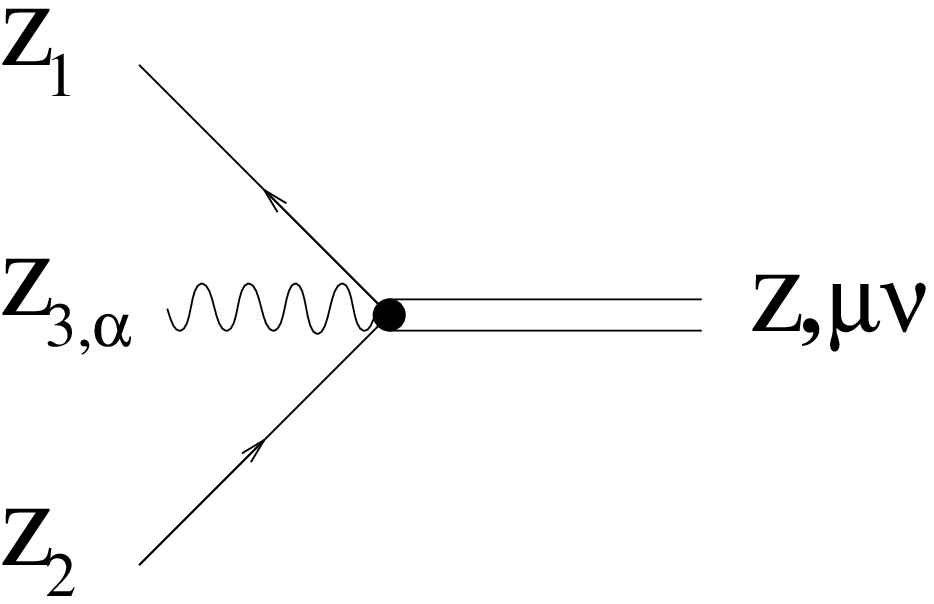}
             }
\end{figure}

\noindent
where the notation is as in (2.9-13). Observe that there is no similar vertex 
involving two scalar legs (as opposed to the two fermion legs we have) as 
such vertex would give diagrams that do not contribute to the abnormal 
parity part of the correlator we are computing.

When this vertex is considered, one finds that there are 7 new diagrams 
that must be included in our calculation, the ones presented in Figure 2. 
There is 1 primitive diagram in Figure 2(a). There are 2 other primitive 
diagrams which only correspond to 1 independent calculation, the one 
depicted in Figure 2(b). Then we have energy-momentum insertion vertex 
corrections. These are Figure 2(c) and Figure 2(d), 2 independent 
calculations corresponding to 4 diagrams due to reflection symmetry. We 
shall now proceed to evaluate these diagrams.

We start with the primitive diagram in Figure 2(a), 
$P'^{(1)}_{\alpha, \mu\nu, \rho\sigma}(z,y,x)$. This diagram is easily 
evaluated as it involves no integrations. Recall that we have to use (3.18) 
alone, whenever one encounters a virtual photon. Setting $z=0$ and 
performing the conformal inversion, the amplitude 
$P'^{(1)}_{\alpha, \mu\nu, \rho\sigma}(0,y,x)$ becomes,
$$
P'^{(1)}_{\alpha, \mu\nu, \rho\sigma}(0,y,x)
=-{igf^2k^2\over64\pi^8}\,{y'}^8{x'}^8\,
J_{\bar{\mu}(\mu}(y')\,J_{\nu)\bar{\nu}}(y')\,J_{\bar{\rho}(\rho}(x')
\,J_{\sigma)\bar{\sigma}}(x')\,
\varepsilon_{\alpha\kappa\bar{\mu}\bar{\rho}}
\,{(x'-y')_\kappa\over(x'-y')^6}\,J_{\bar{\nu}\bar{\sigma}}(x'-y'). 
\eqno(3.24)
$$

Unlike all the preceeding calculations, this amplitude involves no 
derivatives. This is certainly to be expected due to the nature of 
vertex (3.23). However, one can manipulate (3.24) in order to write it as 
the second derivative of a tensor involving the structures (3.6) 
and (3.7) alone. After some calculations, one can show that 
(3.24) can be re-written as (including the factor of 2 for opposite directions 
of fermion charge flow):
$$
P^{(1)}_{\alpha, \mu\nu, \rho\sigma}(0,y,x)\,
=\,{igf^2k^2\over128\pi^8}\,{y'}^8{x'}^8\,
J_{\bar{\mu}(\mu}(y')\,J_{\nu)\bar{\nu}}(y')\,J_{\bar{\rho}(\rho}(x')
\,J_{\sigma)\bar{\sigma}}(x')\cdot
$$
$$
\cdot\Bigl\{\varepsilon_{\alpha\kappa{\bar{\mu}}{\bar{\rho}}}\,
{1\over(x'-y')^2}\,(
{\overrightarrow{\partial}\over\partial {x'}_{\bar{\sigma}}}-
{\overleftarrow{\partial}\over\partial {x'}_{\bar{\sigma}}})\,
{(x'-y')_\kappa\over(x'-y')^2}\,(
{\overrightarrow{\partial}\over\partial {y'}_{\bar{\nu}}}-
{\overleftarrow{\partial}\over\partial {y'}_{\bar{\nu}}})\Bigr\}, 
\eqno(3.25)
$$
and this tensor structure is precisely the same as the one in the 
non-planar diagrams (3.12). This should not come as a surprise as this 
diagram -- like the non-planar ones -- is a primitive.

Let us proceed with the primitive diagram in Figure 2(b), 
$P'^{(2)}_{\alpha, \mu\nu, \rho\sigma}(z,y,x)$. This diagram involves only 
one integration, therefore being different from the ones we previously 
calculated (which either involved two or none integrations). Upon setting 
$z=0$ the amplitude for $P'^{(2)}_{\alpha, \mu\nu, \rho\sigma}(0,y,x)$ is,
$$
P'^{(2)}_{\alpha, \mu\nu, \rho\sigma}(0,y,x)\,=
\,-{igf^2k^2\over128\pi^6}\int {d^4u\over(u-y)^2}\,
J_{\tau\beta}(u-y)\cdot
$$
$$
\cdot{\bf Tr}\,\gamma_\beta\,{{u\sh}\over u^4}\,\gamma_\alpha\gamma_5\,
{{y\sh}\over y^4}\,(\gamma_{(\mu}\delta_{\nu)\tau}-
{1\over2}\,\delta_{\mu\nu}\gamma_\tau)\,S(y-x)\,
\gamma_{(\rho}\delta_{\sigma)\pi}\,(
{\overrightarrow{\partial}\over\partial x_\pi}-
{\overleftarrow{\partial}\over\partial x_\pi})\,S(x-u),	\eqno(3.26)
$$
and performing the conformal inversion one obtains,
$$
P'^{(2)}_{\alpha, \mu\nu, \rho\sigma}(0,y,x)\,=
\,{igf^2k^2\over128\pi^6}\,{y'}^8{x'}^8\,
J_{\bar{\mu}(\mu}(y')\,J_{\nu)\bar{\nu}}(y')\,J_{\bar{\rho}(\rho}(x')
\,J_{\sigma)\bar{\sigma}}(x')\cdot
$$
$$
\cdot\int {d^4{u'}\over(u'-y')^2}\,J_{\bar{\nu}\beta}(u'-y')\,
{\bf Tr}\,\gamma_\beta\,\gamma_\alpha\,\gamma_5\,
\gamma_{\bar{\mu}}\,S(y'-x')\gamma_{\bar{\rho}}\,(
{\overrightarrow{\partial}\over\partial {x'}_{\bar{\sigma}}}-
{\overleftarrow{\partial}\over\partial {x'}_{\bar{\sigma}}})\,
S(x'-u').	\eqno(3.27)
$$

We are left with one integration to perform. However, one should note that 
there is only one differential operator in (3.27), and if we are to obtain 
a final result for this diagram which involves the tensor structures (3.6) 
and (3.7) we shall have to manipulate (3.27) in order to re-write it in such a 
way that it involves two differential operators. This is analogous to the 
situation we faced from (3.24) to (3.25). After integrating and performing 
some calculations, one obtains:
$$
P'^{(2)}_{\alpha, \mu\nu, \rho\sigma}(0,y,x)\,=
\,{igf^2k^2\over128\pi^8}\,{y'}^8{x'}^8\,J_{\bar{\mu}(\mu}(y')\,
J_{\nu)\bar{\nu}}(y')\,J_{\bar{\rho}(\rho}(x')\,J_{\sigma)\bar{\sigma}}(x')\,
\varepsilon_{\alpha\kappa\bar{\mu}\bar{\rho}}\cdot
$$
$$
\cdot\Bigl\{{1\over2}\,{\partial^2\over\partial y'_{\bar{\nu}}\partial 
x'_{\bar{\sigma}}}\,{(x'-y')_\kappa\over(x'-y')^4}\,+\,{1\over(x'-y')^2}\,(
{\overrightarrow{\partial}\over\partial {x'}_{\bar{\sigma}}}-
{\overleftarrow{\partial}\over\partial {x'}_{\bar{\sigma}}})\,
{(x'-y')_\kappa\over(x'-y')^2}\,(
{\overrightarrow{\partial}\over\partial {y'}_{\bar{\nu}}}-
{\overleftarrow{\partial}\over\partial {y'}_{\bar{\nu}}})\Bigr\}. 
\eqno(3.28)
$$

It is interesting to observe that this tensor structure is a linear 
combination of both (3.6) and (3.7). Also, one must now include in this 
result a factor of 2 for opposite directions of fermion charge flow and 
another factor of 2 associated to the 2 distinctive diagrams connected 
through reflection symmetry.

Finally, we can add the 3 diagrams in Figures 2(a) and 2(b), in order to 
obtain the primitive planar contribution to the two loop correlator,
$$
P_{\alpha, \mu\nu, \rho\sigma}(0,y,x)\,=
\,{igf^2k^2\over32\pi^8}\,{y'}^8{x'}^8\,J_{\bar{\mu}(\mu}(y')\,
J_{\nu)\bar{\nu}}(y')\,J_{\bar{\rho}(\rho}(x')\,J_{\sigma)\bar{\sigma}}(x')\,
\varepsilon_{\alpha\kappa\bar{\mu}\bar{\rho}}\cdot
$$
$$
\cdot\Bigl\{{1\over2}\,{\partial^2\over\partial y'_{\bar{\nu}}\partial 
x'_{\bar{\sigma}}}\,{(x'-y')_\kappa\over(x'-y')^4}\,+\,{5\over4}
\,{1\over(x'-y')^2}\,(
{\overrightarrow{\partial}\over\partial {x'}_{\bar{\sigma}}}-
{\overleftarrow{\partial}\over\partial {x'}_{\bar{\sigma}}})\,
{(x'-y')_\kappa\over(x'-y')^2}\,(
{\overrightarrow{\partial}\over\partial {y'}_{\bar{\nu}}}-
{\overleftarrow{\partial}\over\partial {y'}_{\bar{\nu}}})\Bigr\}. 
\eqno(3.29)
$$

\subsection{Diagrams in Figures 1(i), 1(j), 1(k), 1(p), 2(c) and 2(d)}

Next, we proceed with the evaluation of the contributions coming from 
the vertex correction diagrams at the energy-momentum tensor insertions. 
These amplitudes are presented in Figure 1(i): 
$V^{(1)}_{\alpha, \mu\nu, \rho\sigma}(z,y,x)$, 
Figure 1(j): $V^{(2)}_{\alpha, \mu\nu, \rho\sigma}(z,y,x)$, Figure 1(k): 
$V^{(3)}_{\alpha, \mu\nu, \rho\sigma}(z,y,x)$, Figure 1(p): 
$V^{(4)}_{\alpha, \mu\nu, \rho\sigma}$ $(z,y,x)$, and also Figure 2(c): 
$V^{(5)}_{\alpha, \mu\nu, \rho\sigma}(z,y,x)$, and Figure 2(d): 
$V^{(6)}_{\alpha, \mu\nu, \rho\sigma}(z,y,x)$. 
Using the previous treatment with the axial insertion vertex we should 
insert the photon propagator in our calculation in order to guarantee 
finiteness of the energy-momentum insertion vertex. The amplitudes 
$V^{(5)}$ and $V^{(6)}$ have a different structure from the other vertex 
diagrams as they are semi-local. Moreover, there is a possible local term, 
$V^{(4)}$, which is left ambiguous by the Feynman rules. This is analogous 
to the situation we faced when dealing with the self-energy diagrams. As 
this local term cannot be evaluated by Feynman rules it will be solely 
determined from the Ward identity. The role it plays is one of regularizing a 
divergence.

The sum of these amplitudes becomes finite in the finite gauge, which also 
guaranteed the finiteness of the axial insertion vertex. The finite part of 
the vertex subgraphs is a traceless tensor with respect to all three 
included indices, and can be written as,
$$
V^{(finite)}_{\mu\nu}(y,v,u)
={3k(f^2-{1 \over 2} g^2)\over2\pi^4}\,{1\over(y-u)^6}\,\gamma_\kappa\,\Big( 
{(y-u)_\mu\,(y-u)_\nu\,(y-u)_\kappa\over(y-u)^2}\,-\,{1\over 6}\,
\delta_{(\mu\nu}\,(y-u)_{\kappa)}\Big). 	\eqno(3.30)
$$
In addition, it would be thoughtful to check the Ward identity connected 
to these vertices. For that we express the tensor on the RHS 
 of (3.30) in terms of the regularized traceless structure of derivatives,
$$
\Big({(y-u)_\mu\,(y-u)_\nu\,(y-u)_\kappa\over(y-u)^2}-{1\over6}\,
\delta_{(\mu\nu}\,(y-u)_{\kappa)}\Big)\,=\,
-\,{1\over48}\,{\partial\over\partial y_\kappa}\,\Big({\partial\over
\partial y_\mu}\,{\partial\over\partial y_\nu}-{1\over4}\,
\delta_{\mu\nu}\,\Box\,\Big)\,{1\over(y-u)^2}.	\eqno(3.31)
$$
With this expression it is easy to derive that the following Ward identity 
is satisfied for the energy-momentum insertion vertex,
$$
{\partial\over\partial y^{\mu}}\int d^4v \sum_{i=1}^6 V^{(i)}_{\mu\nu}(y,v,u)=
-k\,\partial_{\nu}\,\Sigma(y-u),	\eqno(3.32)
$$
where $V^{(i)}_{\mu\nu}$ is the vertex subgraph in the diagram 
$V^{(i)}_{\alpha, \mu\nu, \rho\sigma}$.

This Ward identity is essential for the calculation of the amplitude of 
Figure 1(i). It suggests that it will give a divergent result, and only when 
we add together the diagrams in Figures 1(i), 1(j), 1(k), 1(p), 2(c) and 2(d) 
we shall obtain a finite answer. After summing up the two possible directions 
of the Higgs field and setting $z=0$ due to translation invariance of the 
amplitude, $V^{(1)}_{\alpha, \mu\nu, \rho\sigma}(0,y,x)$ is written as,
$$
V^{(1)}_{\alpha, \mu\nu, \rho\sigma}(0,y,x)=
-{igf^2k^2\over8\pi^4}\int d^4u\,d^4v\,
\Delta(u-v)\cdot
$$
$$
\cdot{\bf Tr}\,
S(v-y)\gamma_{(\mu}\delta_{\nu)\beta}\,(
{\overrightarrow{\partial}\over\partial y_\beta}-
{\overleftarrow{\partial}\over\partial y_\beta})\,S(y-u)S(u-x)
\gamma_{(\rho}\delta_{\sigma)\pi}\,(
{\overrightarrow{\partial}\over\partial x_\pi}-
{\overleftarrow{\partial}\over\partial x_\pi})\,	
{{x\sh}\over x^4}\,\gamma_\alpha\gamma_5\,
{{v\sh}\over v^4}.	\eqno(3.33)
$$
Using the conformal properties of the theory we can perform the usual 
inversion in the spatial variables which results to,
$$
V^{(1)}_{\alpha, \mu\nu, \rho\sigma}(0,y,x)=
{igf^2k^2\over8\pi^4}\,{y'}^8{x'}^8\,
J_{\bar{\mu}(\mu}(y')\,J_{\nu)\bar{\nu}}(y')\,J_{\bar{\rho}(\rho}(x')
\,J_{\sigma)\bar{\sigma}}(x')
\int d^4{u'}\,d^4{v'}\,\Delta(v'-u')\cdot
$$
$$
\cdot{\bf Tr}\,\gamma_5\,
S(v'-y')\gamma_{\bar{\mu}}\,(
{\overrightarrow{\partial}\over\partial {y'}_{\bar{\nu}}}-
{\overleftarrow{\partial}\over\partial {y'}_{\bar{\nu}}})\,
S(y'-u')S(u'-x')\gamma_{\bar{\rho}}\,
{\overleftarrow{\partial}\over\partial {x'}_{\bar{\sigma}}}
\,\gamma_\alpha.	\eqno(3.34)
$$
In order to proceed with the $u$ and $v$ integrations we should also add 
the diagrams of Figures 1(j), 1(k), 1(p), 2(c) and 2(d) to guarantee 
finiteness of the contribution from the energy-momentum insertion vertex. 
Note that all these diagrams do not contribute with a finite part for the 
order ${\cal O}(gf^2k^2)$ we are interested in, but merely make the diagram 
1(i) finite. This will make the integrand have a traceless form, as given 
in the Appendix. After some manipulations we deduce that,
$$
V''_{\alpha, \mu\nu, \rho\sigma}(0,y,x)
={igf^2k^2\over256\pi^8}\,{y'}^8{x'}^8\,
J_{\bar{\mu}(\mu}(y')\,J_{\nu)\bar{\nu}}(y')\,J_{\bar{\rho}(\rho}(x')
\,J_{\sigma)\bar{\sigma}}(x')\,
\varepsilon_{\alpha\kappa\bar{\mu}\bar{\rho}}\,
{\partial^2\over\partial y'_{\bar{\nu}}\partial x'_{\bar{\sigma}}}
\,{(x'-y')_\kappa\over(x'-y')^4},	\eqno(3.35)
$$
which is proportional to the triangle structure. Taking into account the 2 
fermionic directions and doubling our answer for the two distinctive 
diagrams connected with reflection symmetry we obtain finally:
$$
V'_{\alpha, \mu\nu, \rho\sigma}(0,y,x)={2f^2\over32\pi^2}\,
B_{\alpha, \mu\nu, \rho\sigma}(0,y,x),	\eqno(3.36)
$$
which is similar to what we got in (3.22). So, we can add all the diagrams 
that represent vertex corrections (both at axial and energy-momentum 
insertions). The overall result of the vertex corrections contribution to 
the two loop correlator is,
$$
V_{\alpha, \mu\nu, \rho\sigma}(0,y,x)={3f^2\over32\pi^2}\,
B_{\alpha, \mu\nu, \rho\sigma}(0,y,x).	\eqno(3.37)
$$

\subsection{Diagrams in Figures 1(l), 1(m) and 1(n)}

Finally, we would like to mention the diagrams that are zero. That these 
diagrams vanish can be easily seen either from the fact that the fermion 
trace vanishes or from arguments of Lorentz symmetry. We mention these 
diagrams for completeness. They are the following: Figure 1(l),  which are 
3 diagrams that amount to 1 independent calculation, Figure 1(m), which 
is only 1 diagram, and Figure 1(n), which are 2 diagrams that amount to 1 
independent calculation. We have now completed the calculations
for all the 36 diagrams.

\subsection{The Three Point Function}

The next and final step is to add all diagrams together, and find out what is 
the two loop contribution to the three point function at order 
${\cal O}(gf^2k^2)$. Adding the results for all our diagrams we obtain the 
${\cal O}(gf^2k^2)$ two loop contribution to the correlator 
$\langle A_\alpha(z) T_{\mu\nu}(y) T_{\rho\sigma}(x) \rangle$. There are 4 
distinct contributions: the one from the non-planar primitive diagrams, 
$N_{\alpha, \mu\nu, \rho\sigma}$ $(0,y,x)$ in (3.12); the one from the 
self-energy diagrams, $\Sigma_{\alpha, \mu\nu, \rho\sigma}(0,y,x)$ in (3.17); 
the one from the planar primitive diagrams, 
$P_{\alpha, \mu\nu, \rho\sigma}(0,y,x)$ in (3.29); and the one from the 
vertex correction diagrams, $V_{\alpha, \mu\nu, \rho\sigma}(0,y,x)$ in (3.37). 
Adding these 4 structures we finally obtain our result: the three point 
function does not vanish and consists of two independent conformal tensor 
structures,
$$
N_{\alpha, \mu\nu, \rho\sigma}(0,y,x)\,+\,
\Sigma_{\alpha, \mu\nu, \rho\sigma}(0,y,x)\,+\,
P_{\alpha, \mu\nu, \rho\sigma}(0,y,x)\,+\,
V_{\alpha, \mu\nu, \rho\sigma}(0,y,x)\,=
$$
$$
=\,{igf^2k^2\over64\pi^8}\,{y'}^8{x'}^8\,J_{\bar{\mu}(\mu}(y')\,
J_{\nu)\bar{\nu}}(y')\,J_{\bar{\rho}(\rho}(x')\,J_{\sigma)\bar{\sigma}}(x')\,
\varepsilon_{\alpha\kappa\bar{\mu}\bar{\rho}}\cdot
$$
$$
\cdot\Bigl\{\,7\,{\partial^2\over\partial y'_{\bar{\nu}}\partial 
x'_{\bar{\sigma}}}\,{(x'-y')_\kappa\over(x'-y')^4}\,+\,{11\over2}
\,{1\over(x'-y')^2}\,(
{\overrightarrow{\partial}\over\partial {x'}_{\bar{\sigma}}}-
{\overleftarrow{\partial}\over\partial {x'}_{\bar{\sigma}}})\,
{(x'-y')_\kappa\over(x'-y')^2}\,(
{\overrightarrow{\partial}\over\partial {y'}_{\bar{\nu}}}-
{\overleftarrow{\partial}\over\partial {y'}_{\bar{\nu}}})\,\Bigr\}. 
\eqno(3.38)
$$

There is one consistency check that can be performed on this result. Namely, 
under the appropriate changes, one can ask: does it reduce to the result 
obtained in the axial gauge theory case \cite{JF}? In order to reduce 
(3.38) to the gauge theory case of \cite{JF} we first have to discard 
all diagrams involving the vertex (3.23). Then, in the other diagrams, one 
has to erase all the ``graviton derivatives''. Once this is done we are left 
with a unique conformal tensor, {\it i.e.}, we are reducing the structure of 
our theory to the one in \cite{JF}. Finally, taking into consideration that 
the removal of the ``graviton derivatives'' also includes a factor of $-2$ in 
the non-planar contribution (due to the symmetry enhancement of this 
diagrams), one finds that the overall result vanishes just as it did in 
\cite{JF}. This shows that our result is consistent with the calculations 
performed for the gauge axial anomaly.

We can trace back the reason why this radiative correction does not vanish 
(as it {\it does} vanish in the gauge theory case \cite{JF}). This is (3.12) 
and (3.29), the contribution of the primitive diagrams, which is {\it not} a 
multiple of the one loop amplitude. The existence of two different conformal 
tensors in our theory is a result of the dimensionality of the correlator 
$\langle A_\alpha(z) T_{\mu\nu}(y) T_{\rho\sigma}(x) \rangle$.

\vspace{5 mm}
\noindent
{\bf Acknowledgments:} 
We would like to thank Daniel Freedman and Kenneth Johnson for suggesting and 
guiding the above investigation, as well as for comments and reading of the 
manuscript. We would also like to thank Roman Jackiw for comments and reading 
of the manuscript, and Joshua Erlich for helpful remarks. One of us (R.S.) is 
partially supported by the Praxis XXI grant BD-3372/94 (Portugal).

\vfill
\eject

\appendix

\section{Differential Regularization}

In this Appendix we study the differential regularization of the one loop 
triangle diagram associated to the gravitational axial anomaly, Figure 1(a). 
As we have seen in Section II, the amplitude for this diagram is,
$$
{1\over2}igk^2\,{\bf Tr}\,\gamma_\alpha\gamma_5\,S(z-y)\,
\gamma_{(\mu}\delta_{\nu)i}\,(
{\overrightarrow{\partial}\over\partial y_i}-
{\overleftarrow{\partial}\over\partial y_i})\,S(y-x)\,
\gamma_{(\rho}\delta_{\sigma)j}\,(
{\overrightarrow{\partial}\over\partial x_j}-
{\overleftarrow{\partial}\over\partial x_j})\,S(x-z).	\eqno({\rm A}.1)
$$
We can re-write this diagram as a ``generic'' fermion triangle diagram, for 
which the bare amplitude takes the form:
$$
{\bf Tr}\,\gamma_{III}\,S(z-y)\,
\gamma_{II}\,(
{\overrightarrow{\partial}\over\partial y_i}-
{\overleftarrow{\partial}\over\partial y_i})\,S(y-x)\,
\gamma_{I}\,(
{\overrightarrow{\partial}\over\partial x_j}-
{\overleftarrow{\partial}\over\partial x_j})\,S(x-z). 
\eqno({\rm A}.2)
$$
As compared to (A.1), this is a slightly more general form of the amplitude 
we want to regulate. For our present purposes we shall only need to 
concentrate on the regularization of the singular functions present in the 
amplitude, so we may as well just consider (A.2). 

Performing the derivatives, (A.2) can be written as:
$$
{\bf Tr}\,\Bigl\{\,-{\partial^2\over\partial y_i\partial x_j}\Bigl( 
\gamma_{III}\,S(z-y)\,\gamma_{II}\,S(y-x)\,\gamma_{I}\,S(x-z)\Bigr)\,+
$$
$$
+\,2\gamma_{II}\,{\partial\over\partial y_i}\,S(y-x)\,\gamma_{I}\,
{\partial\over\partial x_j}\,S(x-z)\,\gamma_{III}\,S(z-y)\,+
$$
$$
+\,2\gamma_{III}\,{\partial\over\partial y_i}\,S(z-y)\,\gamma_{II}\,
{\partial\over\partial x_j}\,S(y-x)\,\gamma_{I}\,S(x-z)\,\Bigr\}\,=
$$
$$
=\,-{1\over(4\pi^2)^3}\,{\bf Tr}\,
[\gamma_{III}\,\gamma_a\,\gamma_{II}\,\gamma_b\,\gamma_{I}\,\gamma_c]\,
\Bigl\{\,-{\partial^2\over\partial y_i\partial x_j}\Bigl(\,
{\partial\over\partial z_a}\,{1\over(z-y)^2}\,
{\partial\over\partial y_b}\,{1\over(y-x)^2}\,
{\partial\over\partial x_c}\,{1\over(x-z)^2}\,\Bigr)\Bigr\}\,-
$$
$$
-{2\over(4\pi^2)^3}\,{\bf Tr}\,
[\gamma_{II}\,\gamma_a\,\gamma_{I}\,\gamma_b\,\gamma_{III}\,\gamma_c]\,
\Bigl\{\,
{\partial\over\partial y_i}\,
{\partial\over\partial y_a}\,{1\over(y-x)^2}\,
{\partial\over\partial x_j}\,
{\partial\over\partial x_b}\,{1\over(x-z)^2}\,
{\partial\over\partial z_c}\,{1\over(z-y)^2}\,\Bigr\}\,-
$$
$$
-{2\over(4\pi^2)^3}\,{\bf Tr}\,
[\gamma_{III}\,\gamma_a\,\gamma_{II}\,\gamma_b\,\gamma_{I}\,\gamma_c]\,
\Bigl\{\,
{\partial\over\partial y_i}\,
{\partial\over\partial z_a}\,{1\over(z-y)^2}\,
{\partial\over\partial x_j}\,
{\partial\over\partial y_b}\,{1\over(y-x)^2}\,
{\partial\over\partial x_c}\,{1\over(x-z)^2}\,\Bigr\}.	\eqno({\rm A}.3)
$$

With $x'=x-z$, $y'=y-z$, $\bar{a}=c$, $\bar{b}=a$ and $\bar{c}=b$, we can 
re-write the first term in the previous result as the singular function,
$$
{t^{(1)}_{ij}}_{\bar{a}\bar{b}\bar{c}}(x',y')\,=
\,{\partial^2\over\partial y'_i\partial x'_j} \Bigl(\,
{\partial\over\partial x'_{\bar{a}}}\,{1\over x'^2}\,
{\partial\over\partial y'_{\bar{b}}}\,{1\over y'^2}\,
{\partial\over\partial x'_{\bar{c}}}\,{1\over(x'-y')^2}\,\Bigr), 
\eqno({\rm A}.4)
$$
though in the following we shall drop primes and bars, and simply use the 
notation $(x,y)$ and $\{a,b,c\}$. By a similar re-naming of variables one 
can likewise manipulate the second and third terms in (A.3) to obtain 
-- in both cases -- the singular function,
$$
{t^{(2)}_{ij}}_{abc}(x,y)\,=\,{\partial\over\partial x_j}\,
{\partial\over\partial x_a}\,{1\over x^2}\,
{\partial\over\partial y_i}\,
{\partial\over\partial y_b}\,{1\over y^2}\,
{\partial\over\partial x_c}\,{1\over(x-y)^2}.	\eqno({\rm A}.5)
$$
Observe that (A.5) is independent of (A.4), so that in (A.3) we have 
two independent singular functions. For particular choices of the vertex 
gamma matrices, (A.2) describes the anomalous three point function 
(explicitly studied in the paper) as well as other physically interesting 
amplitudes.

We need to ``pull out'' derivatives and regularize both 
${t^{(1)}_{ij}}_{abc}(x,y)$ and ${t^{(2)}_{ij}}_{abc}(x,y)$. Starting with 
(A.4), one easily observes that this can be written as,
$$
{t^{(1)}_{ij}}_{abc}(x,y)\,=\,{\partial^2\over\partial y_i\partial x_j}
\,t_{abc}(x,y), 	\eqno({\rm A}.6)
$$
where $t_{abc}(x,y)$ is the singular function which is present in the 
bare amplitude for the gauge axial anomaly, and has been 
previously considered in \cite{kj2}. In particular, it was shown in this 
reference how to regularize this singular function. Two derivatives are 
required to control the linear divergence arising from the singularity at 
$x\sim y\sim 0$. Making manifest the $x\leftrightarrow y$, 
$a\leftrightarrow b$ antisymmetry of $t_{abc}(x,y)$ one can write,
$$
t_{abc}(x,y)\,=\,F_{abc}(x,y)\,+\,S_{abc}(x,y), 	\eqno({\rm A}.7)
$$
where the function $F_{abc}(x,y)$ has finite Fourier transform by power 
counting and trace arguments,
$$
F_{abc}(x,y)\,=\,{\partial\over\partial x_a}\,{\partial\over\partial y_b}\,
\Bigl[\,{1\over x^2y^2}\,{\partial\over\partial x_c}\,{1\over(x-y)^2}\,
\Bigr]\,+\,{\partial\over\partial x_a}\Bigl[\,{1\over x^2y^2}\,\Bigl(\,
{\partial^2\over\partial x_b\partial x_c}\,-\,{1\over4}\delta_{bc}\Box\,\Bigr)
\,{1\over(x-y)^2}\Bigl]\,-
$$
$$
-\,{\partial\over\partial y_b}\Bigl[\,{1\over x^2y^2}\,\Bigl(\,
{\partial^2\over\partial x_a\partial x_c}\,-\,{1\over4}\delta_{ac}\Box\,\Bigr)
\,{1\over(x-y)^2}\Bigl]\,-
$$
$$
-\,{1\over x^2y^2}\,\Bigl[\,
{\partial^3\over\partial x_a\partial x_b\partial x_c}\,-\,{1\over6}\,
\Bigl(\,\delta_{ab}{\partial\over\partial x_c}\,+\,
\delta_{bc}{\partial\over\partial x_a}\,
+\,\delta_{ac}{\partial\over\partial x_b}\,\Bigr)\,\Box\,\Bigl]\,
{1\over(x-y)^2}.	\eqno({\rm A}.8)
$$

The term $S_{abc}(x,y)$ contains the ``traces'' subtracted off in (A.8) 
and thus derivatives of $\delta(x-y)$ times $1/x^4$ factors which are 
regulated as is standard in differential regularization, yielding:
$$
S_{abc}(x,y)\,=\,{1\over4}\pi^2\,\Bigl\{\,\Bigl[\,\delta_{bc}
{\partial\over\partial x_a}\,-\,\delta_{ac}{\partial\over\partial y_b}\,
\Bigr]\,\delta(x-y)\,\Box\,{\ln M_1^2x^2 \over x^2}\,-
$$
$$
-\,{1\over3}\Bigl[\,\delta_{bc}\,\Bigl(\,{\partial\over\partial x_a}\,-
\,{\partial\over\partial y_a}\,\Bigr)\,+\,\delta_{ac}\,\Bigl(\,
{\partial\over\partial x_b}\,-\,{\partial\over\partial y_b}\,\Bigr)\,+\,
\delta_{ab}\,\Bigl(\,{\partial\over\partial x_c}\,-
\,{\partial\over\partial y_c}\,\Bigr)\,\Bigr]\,\delta(x-y)\,\Box\,
{\ln M_2^2x^2 \over x^2}\Bigr\}. \eqno({\rm A}.9)
$$

Two different mass scales were used for the two independent trace terms 
in $S_{abc}(x,y)$. Renormalization or symmetry conditions may be used to 
determine the ratio $M_1/M_2$ in particular cases of the triangle 
amplitude.

Expressions (A.8) and (A.9) provide the required regularization of (A.4) 
via expressions (A.6-7). We are thus left with the regularization of 
(A.5), which can be performed in a similar fashion. Again, one can write,
$$
{t^{(2)}_{ij}}_{abc}(x,y)\,=\,{F'_{ij}}_{abc}(x,y)\,+\,{S'_{ij}}_{abc}(x,y). 
\eqno({\rm A}.10)
$$

The function ${F'_{ij}}_{abc}(x,y)$ whose Fourier transform is finite by 
power counting and trace arguments is,
$$
{F'_{ij}}_{abc}(x,y)\,=\,{\partial\over\partial y_i}\,
{\partial\over\partial x_j}\,{\partial\over\partial x_a}\,
{\partial\over\partial y_b}\,\Bigl[\,{1\over x^2y^2}\,
{\partial\over\partial x_c}\,{1\over(x-y)^2}\,\Bigr]\,+\,
{\partial\over\partial y_i}\,{\partial\over\partial x_j}\,\cdot
$$
$$
\cdot\Bigl\{\,{\partial\over\partial x_a}\Bigl[\,{1\over x^2y^2}\,\Bigl(\,
{\partial^2\over\partial x_b\partial x_c}\,-\,{1\over4}\delta_{bc}\Box\,\Bigr)
\,{1\over(x-y)^2}\Bigl]\,-
\,{\partial\over\partial y_b}\Bigl[\,{1\over x^2y^2}\,\Bigl(\,
{\partial^2\over\partial x_a\partial x_c}\,-\,{1\over4}\delta_{ac}\Box\,\Bigr)
\,{1\over(x-y)^2}\Bigl]\,-
$$
$$
-\,{1\over x^2y^2}\,\Bigl[\,
{\partial^3\over\partial x_a\partial x_b\partial x_c}\,-\,{1\over6}\,
\Bigl(\,\delta_{ab}{\partial\over\partial x_c}\,+\,
\delta_{bc}{\partial\over\partial x_a}\,
+\,\delta_{ac}{\partial\over\partial x_b}\,\Bigr)\,\Box\,\Bigl]\,
{1\over(x-y)^2}\,\Bigr\}\,+
$$
$$
+\,{\partial\over\partial y_i}\,{\partial\over\partial x_a}\,\Bigl\{\,
{\partial\over\partial y_b}\Bigl[\,{1\over x^2y^2}\,\Bigl(\,
{\partial^2\over\partial x_j\partial x_c}\,-\,{1\over4}\delta_{jc}\Box\,\Bigr)
\,{1\over(x-y)^2}\Bigl]\,-
$$
$$
-\,{1\over x^2y^2}\,\Bigl[\,
{\partial^3\over\partial x_b\partial x_j\partial x_c}\,-\,{1\over6}\,
\Bigl(\,\delta_{bj}{\partial\over\partial x_c}\,+\,
\delta_{jc}{\partial\over\partial x_b}\,
+\,\delta_{bc}{\partial\over\partial x_j}\,\Bigr)\,\Box\,\Bigl]\,
{1\over(x-y)^2}\,\Bigr\}\,-
$$
$$
-\,{\partial\over\partial y_b}\,{\partial\over\partial x_j}\,\Bigl\{\,
{\partial\over\partial x_a}\Bigl[\,{1\over x^2y^2}\,\Bigl(\,
{\partial^2\over\partial x_i\partial x_c}\,-\,{1\over4}\delta_{ic}\Box\,\Bigr)
\,{1\over(x-y)^2}\Bigl]\,-
$$
$$
-\,{1\over x^2y^2}\,\Bigl[\,
{\partial^3\over\partial x_a\partial x_i\partial x_c}\,-\,{1\over6}\,
\Bigl(\,\delta_{ai}{\partial\over\partial x_c}\,+\,
\delta_{ic}{\partial\over\partial x_a}\,
+\,\delta_{ac}{\partial\over\partial x_i}\,\Bigr)\,\Box\,\Bigl]\,
{1\over(x-y)^2}\,\Bigr\}\,-
$$
$$
-\,{\partial\over\partial y_i}\,{\partial\over\partial y_b}\,\Bigl\{\,
{1\over x^2y^2}\,\Bigl[\,
{\partial^3\over\partial x_a\partial x_j\partial x_c}\,-\,{1\over6}\,
\Bigl(\,\delta_{aj}{\partial\over\partial x_c}\,+\,
\delta_{jc}{\partial\over\partial x_a}\,
+\,\delta_{ac}{\partial\over\partial x_j}\,\Bigr)\,\Box\,\Bigl]\,
{1\over(x-y)^2}\,\Bigr\}\,+
$$
$$
+\,{\partial\over\partial x_j}\,{\partial\over\partial x_a}\,\Bigl\{\,
{1\over x^2y^2}\,\Bigl[\,
{\partial^3\over\partial x_i\partial x_b\partial x_c}\,-\,{1\over6}\,
\Bigl(\,\delta_{ib}{\partial\over\partial x_c}\,+\,
\delta_{bc}{\partial\over\partial x_i}\,
+\,\delta_{ic}{\partial\over\partial x_b}\,\Bigr)\,\Box\,\Bigl]\,
{1\over(x-y)^2}\,\Bigr\}\,-
$$
$$
-\,{\partial\over\partial x_a}\,{\partial\over\partial y_b}\,\Bigl\{\,
{1\over x^2y^2}\,\Bigl[\,
{\partial^3\over\partial x_i\partial x_j\partial x_c}\,-\,{1\over6}\,
\Bigl(\,\delta_{ij}{\partial\over\partial x_c}\,+\,
\delta_{jc}{\partial\over\partial x_i}\,
+\,\delta_{ic}{\partial\over\partial x_j}\,\Bigr)\,\Box\,\Bigl]\,
{1\over(x-y)^2}\,\Bigr\}\,+
$$
$$
+\,{\rm Terms}\ {\rm invovling}\  4^{th} \ {\rm and}\  5^{th} \ {\rm order}\ 
{\rm traceless}\ {\rm derivatives}\ {\rm acting}\ {\rm on}\ {1\over(x-y)^2}. 
\eqno({\rm A}.11)
$$
The fourth and fifth derivatives can be obtained in a straightforward fashion, 
but their explicit form is not relevant and they would occupy a couple 
of pages to write down. Therefore we omit these terms.

Again, the term ${S'_{ij}}_{abc}(x,y)$ contains ``traces'' subtracted from 
(A.11) and so its structure is similar to the one of (A.9), containing the 
usual differential regulated derivatives of $\delta(x-y)$ times $1/x^4$ 
factors. One obtains,
$$
{S'_{ij}}_{abc}(x,y)\,=\,{\partial\over\partial y_i}\,
{\partial\over\partial x_j}\,{1\over4}\pi^2\,\Bigl\{\,\Bigl[\,\delta_{bc}
{\partial\over\partial x_a}\,-\,\delta_{ac}{\partial\over\partial y_b}\,
\Bigr]\,\delta(x-y)\,\Box\,{\ln M_1^2x^2 \over x^2}\,-
$$
$$
-\,{1\over3}\Bigl[\,\delta_{bc}\,\Bigl(\,{\partial\over\partial x_a}\,-
\,{\partial\over\partial y_a}\,\Bigr)\,+\,\delta_{ac}\,\Bigl(\,
{\partial\over\partial x_b}\,-\,{\partial\over\partial y_b}\,\Bigr)\,+\,
\delta_{ab}\,\Bigl(\,{\partial\over\partial x_c}\,-
\,{\partial\over\partial y_c}\,\Bigr)\,\Bigr]\,\delta(x-y)\,\Box\,
{\ln M_2^2x^2 \over x^2}\Bigr\}\,-
$$
$$
-\,{\partial\over\partial y_i}\,{\partial\over\partial x_a}\,
{1\over4}\pi^2\,\Bigl\{\,\delta_{jc}{\partial\over\partial y_b}\,
\delta(x-y)\,\Box\,{\ln M_3^2x^2 \over x^2}\,+\,
{1\over3}\Bigl[\,\delta_{bc}\,\Bigl(\,{\partial\over\partial x_j}\,-
\,{\partial\over\partial y_j}\,\Bigr)\,+\,\delta_{jc}\,\Bigl(\,
{\partial\over\partial x_b}\,-\,{\partial\over\partial y_b}\,\Bigr)\,+
$$
$$
+\,\delta_{jb}\,\Bigl(\,{\partial\over\partial x_c}\,-
\,{\partial\over\partial y_c}\,\Bigr)\,\Bigr]\,\delta(x-y)\,\Box\,
{\ln M_4^2x^2 \over x^2}\Bigr\}\,-\,
{\partial\over\partial y_b}\,{\partial\over\partial x_j}\,
{1\over4}\pi^2\,\Bigl\{\,\delta_{ic}{\partial\over\partial x_a}\,
\delta(x-y)\,\Box\,{\ln M_3^2x^2 \over x^2}\,-
$$
$$
-\,{1\over3}\Bigl[\,\delta_{ic}\,\Bigl(\,{\partial\over\partial x_a}\,-
\,{\partial\over\partial y_a}\,\Bigr)\,+\,\delta_{ac}\,\Bigl(\,
{\partial\over\partial x_i}\,-\,{\partial\over\partial y_i}\,\Bigr)\,+\,
\delta_{ai}\,\Bigl(\,{\partial\over\partial x_c}\,-
\,{\partial\over\partial y_c}\,\Bigr)\,\Bigr]\,\delta(x-y)\,\Box\,
{\ln M_4^2x^2 \over x^2}\Bigr\}\,-
$$
$$
-\,{\partial\over\partial y_i}\,{\partial\over\partial y_b}\,
{\pi^2\over12}\,\Bigl\{
\Bigl[\,\delta_{jc}\,\Bigl({\partial\over\partial x_a}\,-
\,{\partial\over\partial y_a}\Bigr)+\delta_{ac}\,\Bigl(
{\partial\over\partial x_j}\,-\,{\partial\over\partial y_j}\Bigr)+
\delta_{aj}\,\Bigl({\partial\over\partial x_c}\,-
\,{\partial\over\partial y_c}\Bigr)\Bigr]\,\delta(x-y)\,\Box\,
{\ln M_5^2x^2 \over x^2}\Bigr\}\,+
$$
$$
+\,{\partial\over\partial x_j}\,{\partial\over\partial x_a}\,
{\pi^2\over12}\,\Bigl\{
\Bigl[\,\delta_{bc}\,\Bigl({\partial\over\partial x_i}\,-
\,{\partial\over\partial y_i}\Bigr)+\delta_{ic}\,\Bigl(
{\partial\over\partial x_b}\,-\,{\partial\over\partial y_b}\Bigr)+
\delta_{ib}\,\Bigl({\partial\over\partial x_c}\,-
\,{\partial\over\partial y_c}\Bigr)\Bigr]\,\delta(x-y)\,\Box\,
{\ln M_5^2x^2 \over x^2}\Bigr\}\,-
$$
$$
-\,{\partial\over\partial x_a}\,{\partial\over\partial y_b}\,
{\pi^2\over12}\,\Bigl\{
\Bigl[\,\delta_{jc}\,\Bigl({\partial\over\partial x_i}\,-
\,{\partial\over\partial y_i}\Bigr)+\delta_{ic}\,\Bigl(
{\partial\over\partial x_j}\,-\,{\partial\over\partial y_j}\Bigr)+
\delta_{ij}\,\Bigl({\partial\over\partial x_c}\,-
\,{\partial\over\partial y_c}\Bigr)\Bigr]\,\delta(x-y)\,\Box\,
{\ln M_6^2x^2 \over x^2}\Bigr\}\,+
$$
$$
+\,{\rm Traces}\ {\rm subtracted}\ {\rm from}\ ({\rm A}.11)\ {\rm in}\ 
{\rm the}\  4^{th} \ {\rm and}\  5^{th} \ {\rm order}\ {\rm derivatives},
$$
$$
{\rm and}\ {\rm their}\ {\rm respective}\ {\rm mass}\ {\rm scales}.
\eqno({\rm A}.12)
$$

One can see that several different mass scales were introduced for the 
several independent trace terms in ${S'_{ij}}_{abc}(x,y)$. Again, as 
was mentioned for (A.9), renormalization or symmetry conditions may be used 
to determine the ratios between the mass scales in particular cases 
of the triangle amplitude.

\vfill
\eject

\section{Convolution Integrals}

In this Appendix we list the convolution integrals that are required 
in order to perform the two loop calculation \cite{rosner}. They are 
inclosed to make this work self contained for the reader who wishes 
to reproduce our result. The table of convolution integrals is (defining 
$\Delta\equiv x-y$, and using the cutoff $\Lambda$):
$$
\int {d^4v\over v^2\,(v-x)^2}\,=\,-\pi^2\,\ln {x^2\over \Lambda^2}, 
\eqno({\rm B}.1)
$$

$$
\int {(v-x)_\rho \over v^2\,(v-x)^4}\,d^4v\,=\,-\pi^2\,{x_\rho 
\over x^2},	\eqno({\rm B}.2)
$$

$$
\int {(v-x)_\rho(v-y)_\sigma \over (v-x)^4\,(v-y)^4}\,d^4v\,=\,
{\pi^2 \over 2\Delta^2}\,(\delta_{\rho\sigma}-2{\Delta_\rho\Delta_\sigma 
\over \Delta^2}),	\eqno({\rm B}.3)
$$

$$
\int {(v_\rho v_\sigma -{1\over4}\delta_{\rho\sigma}v^2) \over 
v^4\,(v-x)^2}\,d^4v\,=\,{\pi^2\over2x^2}\,(x_\rho x_\sigma-
{1\over4}\delta_{\rho\sigma}x^2),	\eqno({\rm B}.4)
$$

$$
\int {((v-x)_\rho(v-x)_\sigma-{1\over4}\delta_{\rho\sigma}
(v-x)^2)(v-y)_\lambda \over (v-x)^4\,(v-y)^4}\,d^4v=
$$
$$
=-{\pi^2 \over 
4\Delta^2}\,(\delta_{\rho\lambda}\Delta_\sigma+\delta_{\sigma\lambda}
\Delta_\rho-2{\Delta_\rho\Delta_\sigma\Delta_\lambda\over
\Delta^2}),	\eqno({\rm B}.5)
$$

$$
\int {(v_\rho v_\sigma -{1\over4}\delta_{\rho\sigma}v^2) \over 
v^6\,(v-x)^2}\,d^4v\,=\,{\pi^2\over2x^4}\,(x_\rho x_\sigma-
{1\over4}\delta_{\rho\sigma}x^2),	\eqno({\rm B}.6)
$$

$$
\int {((v-x)_\rho(v-x)_\sigma-{1\over4}\delta_{\rho\sigma}
(v-x)^2)(v-y)_\lambda \over (v-x)^6\,(v-y)^4}\,d^4v\,=
$$
$$
=-{\pi^2 \over 
4\Delta^4}\,(\delta_{\rho\lambda}\Delta_\sigma+\delta_{\sigma\lambda}
\Delta_\rho+{1\over2}\delta_{\rho\sigma}\Delta_\lambda-
4{\Delta_\rho\Delta_\sigma\Delta_\lambda\over
\Delta^2}),	\eqno({\rm B}.7)
$$

$$
\int {v_\alpha(v-x)_\rho(v-y)_\sigma \over (v-x)^4\,(v-y)^4}\,d^4v\,=\,
{\pi^2 \over 4\Delta^2}\,(\delta_{\alpha\rho}\Delta_\sigma-
\delta_{\alpha\sigma}\Delta_{\rho}+(x+y)_\alpha\,[
\delta_{\rho\sigma}-2{\Delta_\rho\Delta_\sigma 
\over \Delta^2}]),	\eqno({\rm B}.8)
$$

$$
\int {v_\alpha(v-y)_\rho \over (v-x)^2\,(v-y)^4}\,d^4v\,=\,{\pi^2\over2
\Delta^2}\,((x+y)_\alpha\Delta_\rho-{1\over4}\delta_{\alpha\rho}
\Delta^2)-{\pi^2\over4}\delta_{\alpha\rho}\ln {\Delta^2\over
\Lambda^2}.	\eqno({\rm B}.9)
$$

\vfill
\eject

\begin{figure}[h]
  \centerline{
    \epsfxsize=7.5in
    \epsfysize=8.75in
    \epsffile{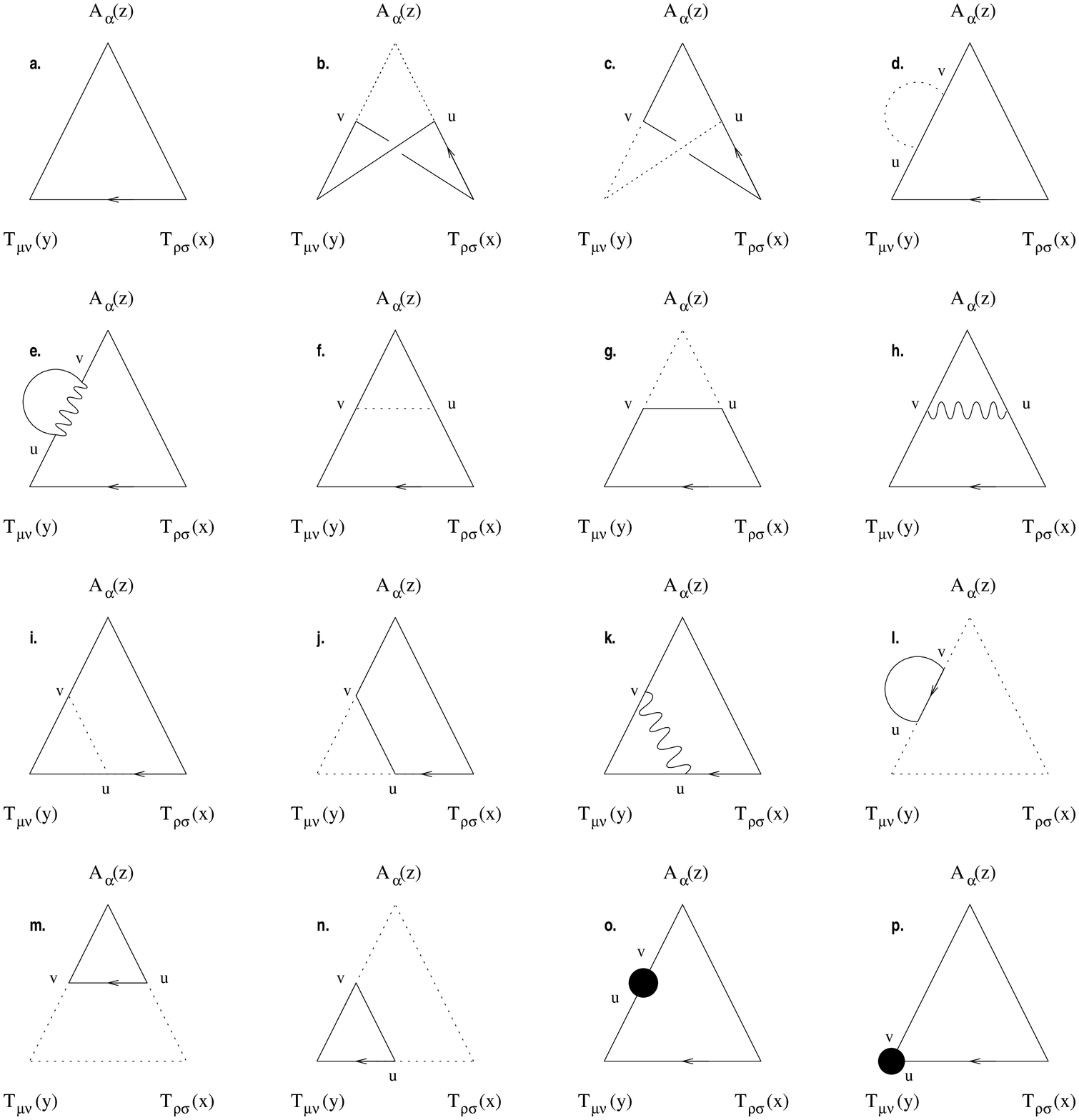}
             }
  \caption[contour]{\label{Fig1}
One and two loop contributions to the anomaly in the Abelian Higgs theory.
}
\end{figure}

\vfill
\eject

\begin{figure}[h]
  \centerline{
    \epsfxsize=3.75in
    \epsfysize=4.3in
    \epsffile{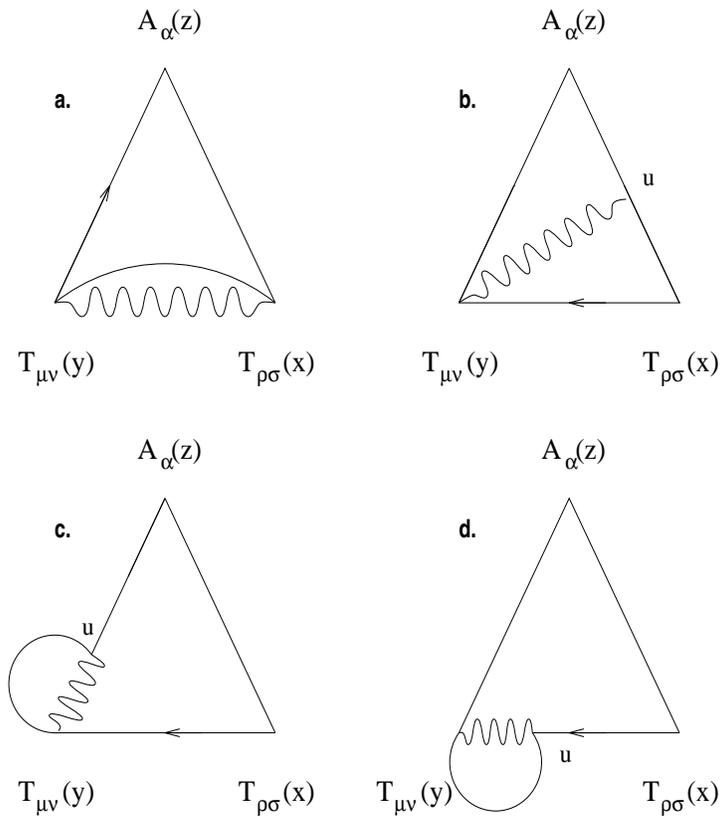}
             }
  \caption[contour]{\label{Fig2}
Two loop contributions to the anomaly involving the four-point vertex.
}
\end{figure}

\vfill
\eject

\end{document}